\documentclass[iop]{emulateapj}

\usepackage{natbib}
\usepackage{longtable}

\newcommand{\HII}{H~{\footnotesize II}} 
\newcommand{\HeII}{He~{\footnotesize II}} 
 
\newcommand{\NII}{[N~{\footnotesize II}]}

\newcommand{\OI}{[O~{\footnotesize I}]} 
\newcommand{\OIII}{[O~{\footnotesize III}]} 
\newcommand{\SII}{[S~{\footnotesize II}]} 

\newcommand{\fluxunits}{erg~s$^{-1}$~cm$^{-2}$}

\newcommand{\ha}{H$\alpha$}
\newcommand{\hbeta}{H$\beta$}

\newcommand{\msun}{M$_\odot$}
\newcommand{\msigma}{$M_{\rm BH} - \sigma_{\star}$}

\shorttitle{Dwarf Galaxies with Active Massive Black Holes}
\shortauthors{Reines et al.}

\begin{document}

\title{Dwarf Galaxies with Optical Signatures of Active Massive Black Holes}

\author{Amy E. Reines\altaffilmark{1}}
\affil{National Radio Astronomy Observatory,
    Charlottesville, VA 22903, USA}
\email{areines@nrao.edu}


\author{Jenny E. Greene}
\affil{Department of Astrophysical Sciences, Princeton University, Princeton, NJ 08544, USA}
    
\and

\author{Marla Geha}
\affil{Department of Astronomy, Yale University, New Haven, CT 06520, USA}

\altaffiltext{1}{Einstein Fellow}

\begin{abstract}

We present a sample of 151 dwarf galaxies ($10^{8.5} \lesssim M_{\star} \lesssim 10^{9.5}$~\msun) that exhibit optical spectroscopic signatures of accreting massive black holes (BHs), increasing the number of known active galaxies in this stellar mass range by more than an order of magnitude.  Utilizing data from the Sloan Digital Sky Survey Data Release 8 and stellar masses from the NASA-Sloan Atlas, we have systematically searched for active BHs in $\sim$25,000 emission-line galaxies with stellar masses comparable to the Magellanic Clouds and redshifts $z<0.055$.   Using the narrow-line \OIII/\hbeta\ versus \NII/\ha\ diagnostic diagram, we find photoionization signatures of BH accretion in 136 galaxies, a small fraction of which also exhibit broad \ha\ emission.  For these broad-line AGN candidates, we estimate BH masses using standard virial techniques and find a range of $10^5 \lesssim M_{\rm BH} \lesssim 10^6$~M$_\odot$ and a median of $M_{\rm BH} \sim 2 \times 10^5$~M$_\odot$.   We also detect broad \ha\ in 15 galaxies that have narrow-line ratios consistent with star-forming galaxies.  Follow-up observations are required to determine if these are true type 1 AGN or if the broad \ha\ is from stellar processes.  The median absolute magnitude of the host galaxies in our active sample is $M_g = -18.1$ mag, which is $\sim$1--2 magnitudes fainter than previous samples of AGN hosts with low-mass BHs.  This work constrains the smallest galaxies that can form a massive BH, with implications for BH feedback in low-mass galaxies and the origin of the first supermassive BH seeds. 

\end{abstract}


\section{Introduction}\label{sec:intro}

Over the past decade we have come to appreciate that nuclear black holes (BHs) with $M_{\rm BH} \sim 10^6-10^9$~M$_\odot$ are a ubiquitous component of massive galaxies in the modern Universe \citep[e.g.,][]{kormendyrichstone1995,kormendyho2013}, yet the origin of the first high-redshift supermassive BH ``seeds'' is far from understood.  Observations of quasars with billion solar-mass BHs at a time when the Universe was less than a Gyr old \citep[e.g.,][]{fanetal2001,mortlocketal2011} demonstrate that supermassive BHs almost certainly started out with masses considerably in excess of normal stellar-mass BHs.  However, we do not know how these initial seed BHs formed in the early Universe, how massive they were originally, or in what types of galaxies they formed.  While direct observations of distant seed BHs and their host galaxies in the infant Universe are unobtainable with current capabilities, nearby dwarf galaxies are within observational reach and can provide important constraints on the formation path, masses and hosts of BH seeds \citep[e.g.,][]{volonteri2010,greene2012}.  The goal of this work is to systematically search for optical AGN signatures in dwarf galaxies ($M_{\star} \lesssim 3 \times 10^9$~\msun), where very few massive BHs\footnote{We use the term ``massive BH" to refer to BHs larger than normal stellar-mass BHs.  BHs with masses in the range $\sim 10^4-10^6$~M$_\odot$ are also sometimes referred to as ``low-mass BHs" (meaning relatively low-mass supermassive BHs) or ``intermediate-mass BHs (IMBHs)."} have hitherto been found.

The growth of supermassive BHs appears to be linked to the evolution of their hosts, with more massive galaxies generally harboring more massive BHs \citep[e.g.,][]{gebhardtetal2000a,ferraresemerritt2000,marconihunt2003,gultekinetal2009,mcconnellma2013}.  Therefore, unlike today's massive galaxies, low-mass dwarf galaxies with relatively quiet merger histories may host BHs with masses not so different from the first seed BHs \citep{bellovaryetal2011}.  Models of BH growth in a cosmological context \citep{volonterietal2008} indicate that if we can determine the BH occupation fraction and mass distribution in present-day dwarf galaxies, we will gain insight into whether seed BHs formed primarily as the end-product of Population III stars \citep[e.g.,][]{brommyoshida2011}, in a direct collapse scenario \citep[e.g.,][]{haehneltrees1993, lodatonatarajan2006, begelmanetal2006}, or in a runaway accretion event at the center of a dense star cluster \citep[e.g.,][]{portegieszwartetal2004, millerdavies2012}.  Studying the scaling relations between BHs and galaxies at low-mass may provide further clues \citep[e.g.,][]{volonterinatarajan2009}.

There are additional reasons to search for and study massive BHs in dwarf galaxies.  Understanding the radiative properties of BHs in this mass range is important for models of the impact of mini-quasars on reionization and the formation of the first galaxies \citep[e.g.,][]{milosetal2009}.  Moreover, the study of BH accretion and star formation in low-mass dwarfs today may prove a good laboratory for understanding their interplay at early times \citep[e.g.,][]{reinesetal2011, jiaetal2011, alexandroffetal2012} and the apparent contribution of obscured AGNs in blue low-mass galaxies to the cosmic X-ray background \citep{xueetal2012}.  Determining the presence of AGN in low-mass galaxies will also help constrain BH feedback and galaxy formation models at all mass scales.  For instance, while BH feedback is often invoked to explain star formation quenching in massive galaxies \citep[e.g.,][]{crotonetal2006,kimmetal2009}, the quenched fraction of central galaxies decreases with stellar mass \citep[][]{wetzeletal2012} and may become zero below a stellar mass near $10^9$~\msun\ \citep{kauffmannetal2003b,gehaetal2012}.  Finally, signatures of tidal disruptions of stars are strongest for low-mass BHs \citep[e.g.,][]{strubbequataert2009,rosswogetal2009,millergultekin2011} and $\sim 10^5$~\msun\ BH mergers have the right frequency to provide a signal for future gravitational wave experiments \citep[e.g.,][]{hughes2002}. 

The most secure method of discovering supermassive BHs uses stellar or gas dynamics to weigh the central BH \citep[see review in][]{kormendy2004}.  However, the gravitational sphere of influence cannot be resolved for low-mass BHs in small galaxies much beyond the Local Group.  For example, $r_{\rm infl} = G M_{\rm BH}/\sigma^2 \sim 0.5$~pc assuming $M_{\rm BH} \sim 10^5~$\msun\ and a stellar velocity dispersion of $\sigma \sim 30$~km s$^{-1}$.  Even with the resolution afforded by the {\it Hubble Space Telescope (HST)} or adaptive optics on large ground-based telescopes, such a small radius of influence can only be resolved out to a distance of $\sim$ 1 Mpc.  Nevertheless, there are limits on dynamical BH masses and a couple of detections in nearby low-mass galaxies.  The Local Group galaxies NGC 205 (a bright early-type dwarf satellite of Andromeda), Ursa Minor, and Fornax (both classical dwarf spheroidals around the Milky Way) all have upper limits of $M_{\rm BH} \lesssim$ a few $\times 10^4$~\msun\ from stellar dynamics \citep{vallurietal2005, loraetal2009, jardelgebhardt2012}.  In contrast, the low-mass elliptical M32 has a BH with $M_{\rm BH} \sim$ a few $\times 10^6$~\msun\ \citep[e.g.,][]{dresslerrichstone1988, vandermareletal1998}.  NGC 404, a low-mass S0 just beyond the Local Group, also has a likely BH with $M_{\rm BH} \sim 5 \times 10^5$~\msun\ derived from molecular hydrogen gas kinematics \citep{sethetal2010}.  A number of very late-type spiral galaxies have dynamical BH mass limits, including M33 with $M_{\rm BH} \lesssim$ a few $\times 10^3$~\msun\ \citep{gebhardtetal2001}.  Outside the Local Group, \citet{barthetal2009} find a conservative upper limit of $M_{\rm BH} \lesssim 3 \times10^6$~\msun\ on the active BH in NGC 3621, while \citet{neumayerwalcher2012} find upper limits of $\sim 10^6$~\msun\ for nine more late-type spirals. 

In more distant systems, and in particular those with low-mass BHs, obtaining dynamical evidence for a BH is not feasible and we must rely on radiative signatures of AGN powered by BH accretion.  The dwarf spiral galaxy NGC 4395 ($M_{\star} \sim 1.3 \times 10^9$~\msun), for example, has clear and unambiguous evidence for a central BH from high-ionization narrow emission lines and broad permitted lines in the optical \citep{filippenkosargent1989}, a compact radio jet \citep{wrobelho2006}, and extreme variability in the X-ray \citep{vaughanetal2005,moranetal2005}.  NGC 4395 is not just a fluke; POX 52 is dwarf elliptical \citep[$M_{\star} \sim 1.2 \times 10^9$~\msun;][]{thorntonetal2008} with an AGN that is very similar in optical properties to NGC 4395 \citep{kunthetal1987,barthetal2004}.  These two dwarf galaxies have BH mass estimates of $M_{\rm BH} \sim 3 \times 10^5$~\msun\ \citep{filippenkoho2003,petersonetal2005,thorntonetal2008}.  

\citet{reinesetal2011} present multi-wavelength evidence for the first example of a massive BH in a dwarf starburst galaxy, Henize 2-10, which has an irregular central morphology and no discernible bulge.  The evidence for an accreting massive BH includes Very Large Array radio and {\it Chandra} hard X-ray point sources at the center of the galaxy and clearly separated from the bright \HII\ regions.  Follow-up Very Long Baseline Interferometry (VLBI) observations reveal parsec-scale non-thermal radio continuum from the precise location of the putative active nucleus \citep{reinesdeller2012} and no star clusters are seen at this location in {\it HST} observations.  The source lies at the center of a $\sim$250-pc-long ionized gas structure that is suggestive of bipolar flow and the position of the central source is consistent with the dynamical center of the galaxy.  \citet{reinesetal2011} estimate a BH mass of log ($M_{\rm BH}/$\msun) = 6.3 $\pm$ 1.1 using the radio--X-ray--$M_{\rm BH}$ fundamental plane \citep{merlonietal2003}, and \citet{kormendyho2013} estimate that Henize 2-10 has a stellar mass of $M_{\star} \sim 1.4 \times 10^9$~\msun\ with an uncertainty of a factor of 3.  To determine the frequency of objects like NGC 4395, Pox 52, and Henize 2-10 requires large surveys for AGN signatures in low-mass galaxies. 

In an effort to find AGNs with {\it low-mass BHs}, \citet{greeneho2004,greeneho2007}, and \citet{dongetal2012} searched systematically through the Sloan Digital Sky Survey \citep[SDSS; ][]{yorketal2000} for broad H$\alpha$, as a signature that an accreting BH is present.  They present samples of $> 200$ galaxies with $M_{\rm BH} \lesssim 10^{6.5}$~\msun\ and median BH masses of $M_{\rm BH} \sim 1 \times 10^{6}$~\msun.  In a complementary search, \citet{barthetal2008a} present a sample of narrow-line AGN in host galaxies with relatively low stellar velocity dispersions, suggesting the presence of low-mass BHs.  However, the vast majority of galaxies in all of these samples are more massive than $M_\star \sim 10^9$~\msun, and thus do not probe the dwarf galaxy regime \citep{barthetal2008a, greeneho2008, jiangetal2011}.  Determining the true space densities of low-mass BHs from the SDSS surveys is very difficult because of the joint bias towards luminous galaxies (for spectroscopic targeting) and luminous AGNs \citep[for spectroscopic identification;][]{greeneho2007b}.  Searches based on mid-infrared \citep[e.g., Spitzer spectroscopy;][]{satyapaletal2007, satyapaletal2008, satyapaletal2009, gouldingetal2010} or X-ray observations \citep{ghoshetal2008, desrochesho2009, galloetal2008, galloetal2010, kamizasaetal2012, milleretal2012,schrammetal2013} have so far covered only small volumes, often with heterogeneous selection.

So far there has been no systematic look at the AGN population in dwarf galaxies.  Previous AGN studies have looked at galaxies with stellar masses typically larger than $\sim 10^{10}$~\msun.  Here we undertake the first systematic search for active massive BHs in dwarf galaxies with stellar masses comparable to or less than the Large Magellanic Cloud (LMC).  Starting with a sample of $\sim 25,000$ emission-line galaxies in the SDSS Data Release 8 (DR8) spectroscopic catalog \citep{aiharaetal2011} with stellar masses $M_\star \lesssim 3 \times 10^9$~\msun\ and redshifts $z < 0.055$, we identify dwarf galaxies with narrow-line photoionization signatures of an accreting massive BH.  In addition, we search for broad \ha\ emission that may indicate gas orbiting in the deep potential well of a massive BH.  Collectively, the galaxies presented here are the smallest and least-massive known to contain massive BHs.


\section{Data and Dwarf Galaxy Sample}\label{sec:sample}

We have selected our sample of dwarf galaxies from the NASA-Sloan Atlas\footnote{http://www.nsatlas.org} (NSA), which in turn is based on the SDSS DR8 spectroscopic catalog \citep{yorketal2000, aiharaetal2011}.  We use the NSA for selecting our parent sample of galaxies and investigating galaxy properties, but analyze the SDSS spectra with our own customized software to search for signatures of BH accretion (Section \ref{sec:analysis}).  The SDSS uses the dedicated 2.5-meter wide-field Sloan Foundation Telescope and a 640-fiber double spectrograph at Apache Point Observatory in New Mexico \citep{gunnetal2006}. The instrumental fiber diameter is 3\arcsec\ and the spectrophotometrically calibrated spectra cover a wavelength range from $3800-9200$~\AA, with a instrumental dispersion of 69 km s$^{-1}$ per pixel. 

The NSA provides a re-analysis of the SDSS imaging and spectroscopic data for all galaxies with redshifts $z < 0.055$.  Photometry is improved over the standard SDSS DR8 photometric catalog as described in \citet{blantonetal2011} and results in a cleaner sample of dwarf galaxies compared to previous catalogs that include many fragmented pieces of extended massive galaxies.  The spectroscopic data is re-analyzed using the methods of \citet{yanblanton2012} and \citet{yan2011}.  We only use the emission line measurements in the NSA for quality cuts in signal-to-noise and equivalent width as described below.  In addition to a variety of galaxy parameters, estimates of stellar masses are provided in the NSA.  Stellar masses are derived from the \texttt{kcorrect} code of \citet{blantonroweis2007}, which fits broadband fluxes using templates based on the stellar population synthesis models of \citet{bruzualcharlot2003} and the nebular emission-line models of \cite{kewleyetal2001}.   Masses are given in units of M$_\odot h^{-2}$ and we have assumed $h=0.73$. 

Since our goal is to search for dwarf galaxies hosting active massive BHs, we first select sources in the NSA with stellar masses $M_{\star} \leq 3 \times 10^9$ M$_\odot$ and obtain 44,594 objects.  Our mass threshold, while somewhat arbitrary, is approximately equal to the stellar mass of the LMC \citep{vandermareletal2002}, which is the most massive dwarf galaxy satellite around the Milky Way.  While we do not apply a minimum stellar mass limit, nearly all of the galaxies in our parent sample have $M_{\star} \gtrsim 10^7$ M$_\odot$ due to the SDSS spectroscopic apparent magnitude limit of $r < 17.7$.  Therefore, while we can probe dwarf galaxies with stellar masses comparable to the Magellanic Clouds, we are not sensitive to galaxies with lower stellar mass such as the Fornax dwarf spheroidal ($2 \times 10^7$ M$_\odot$) around the Milky Way \citep[e.g.,][]{mcconnachie2012, mateo1998}.  We also impose modest requirements on emission line flux measurements reported in the NSA: a signal-to-noise ratio $S/N \geq 3$ and an equivalent width EW $>$ 1 for \ha, \NII\ $\lambda 6584$, and \OIII\ $\lambda 5007$ and $S/N \geq 2$ for \hbeta.  This reduces the number of sources to 25,974.
 
We use stellar mass as our primary selection criterion rather than absolute magnitude since galaxies of a given mass can span a wide range in absolute magnitude due to differences in stellar populations.  Younger, bluer, star-forming galaxies will be brighter than older, redder, quenched galaxies for a given mass \citep[e.g.,][]{belletal2003}.  Additionally, strong emission lines (and even nebular continuum) can boost broad-band fluxes, especially in the SDSS $g$- and $r$-bands \citep[e.g.,][]{reinesetal2010,andersetal2003}.  Therefore, imposing an absolute magnitude cut would preferentially exclude line-emitting galaxies compared to equivalent galaxies that do not have strong emission lines.  The derived stellar masses are less sensitive to these effects since multiple bands are fit and the SED templates of \citet{blantonroweis2007} include emission lines.  However, imposing a stellar mass cut introduces a bias such that blue/star-forming galaxies will be detected over a larger volume than red/quenched galaxies with the same stellar mass \citep[e.g.,][]{gehaetal2012}.

\section{Analysis and Results}\label{sec:analysis}

There are many ways to identify the presence of an accreting supermassive BH \citep[see the review in][]{ho2008}.  In this work, we search for two signatures of active BHs in the optical spectra of galaxies: 1) narrow emission-line ratios indicating photoionization by an accreting BH, and 2) broad H$\alpha$ emission signifying dense gas orbiting a massive BH.  After selecting dwarf emission-line galaxies from the NSA (\S\ref{sec:sample}), we retrieve the SDSS spectra for the entire sample and analyze them with customized software as described below.  All code was developed in the Interactive Data Language (IDL) and our fitting routines make use of the non-linear least squares curve fitting package MPFIT \citep{markwardt2009}.  Individual sources flagged by our automated algorithms are subsequently inspected by eye.

\subsection{Continuum and Absorption Line Subtraction}\label{sec:contsub}

Before we can look for any signature of an AGN, we need to model and remove stellar continuum from the host galaxy, that in almost all cases dominates the total continuum.  The spectra also contain stellar absorption lines.  Balmer absorption is of particular concern since we aim to detect potentially weak broad \ha\ emission from nuclear activity in the galaxies.  Our general approach for modeling the continuum and absorption lines is based on that presented in \citet{tremontietal2004}.  We fit the galaxy spectra with a non-negative linear combination of simple stellar population (SSP) models spanning a range of ages for a given metallicity.  The model templates are the same as those used in \citet{tremontietal2004} and come from the stellar population synthesis code of \citet{bruzualcharlot2003}\footnote{We retrieved the templates from the corresponding website http://www2.iap.fr/users/charlot/bc2003/.}.  The templates include model spectra for 10 different ages (0.005, 0.025, 0.1, 0.29, 0.64, 0.9, 1.4, 2.5, 5, and 11 Gyr) and 3 different metallicities ($Z$ = 0.008, 0.02, 0.05).  Each galaxy spectrum is modeled as a combination of SSP templates with a single metallicity and we select the metallicity yielding the best-fit.  In the fitting process, we allow for reddening from dust using the Galactic extinction curve of \citet{cardellietal1989} and Gaussian smoothing to match the absorption line widths.  

Our algorithm for modeling the continuum and absorption lines in each spectrum is an iterative process.  The SSP models do not contain emission lines and so we begin by masking out pixels 5$\sigma$ above the continuum in the observed spectra to remove both strong and weak lines.  We then apply a redshift-correction using the redshift derived from emission lines provided in the NSA catalog.  A preliminary model is fit to the resulting spectrum and the two are cross-correlated to determine any redshift offset between the absorption and emission lines.  After correcting for the redshift offset, we find the best fit model for each of the 3 model metallicities.  The single-metallicity model with the smallest $\chi^2$ value is selected and used to improve the masking of emission lines.  Pixels more than 5$\sigma$ above the difference spectrum (data $-$ model) are excluded in the last round of fitting.  Again, we find the best-fit model for each metallicity and select the single-metallicity model with the smallest $\chi^2$ for our final model of the continuum and absorption line spectrum.  The vast majority of the galaxy spectra are best fit by the sub-solar metallicity model ($Z$ = 0.008), which is consistent with studies demonstrating that low-mass galaxies generally have low metallicities \citep[e.g.][]{tremontietal2004}.  In the small fraction of cases where absorption lines are either extremely weak or absent, a third-order polynomial is fit to the continuum.  We subtract this model from the data to produce a pure-emission line spectrum.  

\subsection{Emission Line Measurements}\label{sec:linefit}

In order to identify the spectral signatures of active massive BHs, we model the emission lines of interest as Gaussians.  We are especially interested in detecting any possible broad \ha\ component.  Luminous unobscured AGN powered by supermassive BHs in massive galaxies often exhibit bright, broad \ha\ emission that is clearly visible even by eye.  However, the broad \ha\ signatures from AGNs with lower-mass BHs hosted by dwarf galaxies will be less pronounced, generally having narrower widths and lower luminosities than their higher mass counterparts.  Moreover, any potential broad \ha\ feature will be blended with narrow \ha\ emission and the surrounding \NII\ $\lambda\lambda 6548, 6583$ doublet.  Therefore, it is vital to carefully model the narrow emission line profile in order to detect any potential broad \ha\ emission.

We base our model of the narrow emission line profile on the \SII\ $\lambda\lambda 6713, 6731$ doublet, which has been shown to be generally well-matched to the line profiles of the \NII\ $\lambda\lambda 6548, 6583$ doublet and narrow \ha\ \citep{filippenkosargent1988,filippenkosargent1989,hfs1997broad,greeneho2004}.  The \SII\ doublet is first fit with a single Gaussian model with the width of the two lines assumed to be equal and the relative separation between the two lines held fixed by their laboratory wavelengths.  The \SII\ doublet is also fit with a two-component Gaussian model, with the added constraint that the height ratio of the two components must be equal for each line.  If the reduced $\chi^2$ value from the two-component model is at least $10\%$ lower than that of the single Gaussian model, the two-component model is selected.  Nearly all of the galaxies in our dwarf sample only require a single Gaussian to model the \SII\ doublet, with only 15 sources preferring a two-component model (8 of which turn out to be classified as either a BPT AGN or composite, see Section \ref{sec:narrow}).  

Once we have constructed a suitable model of the \SII\ doublet, we use it as a template for fitting the narrow lines in the \ha\ + \NII\ complex.  The relative separations between the centroids of the narrow Gaussian components are held fixed using laboratory wavelengths and the flux of \NII\ $\lambda$6584 to \NII\ $\lambda$6548 is fixed at the theoretical value of 2.96.  We assume the \NII\ lines have the same width (in velocity) as the \SII\ lines, but generously allow the width of the narrow \ha\ component to increase by as much as 25\% for the vast majority of cases in which a single Gaussian is used to model the narrow line profile.  For the small number of sources (15/25974) with two-component Gaussian models for the narrow lines, the profiles are strictly scaled from the \SII\ lines.  The \ha\ + \NII\ complex is fit and the reduced $\chi^2$ computed.  The \ha\ + \NII\ complex is fit a second time, allowing for an additional broad \ha\ component.  If the resulting reduced $\chi^2$ is improved by at least 20\% and the FWHM of the broad \ha\ component is at least 500 km s$^{-1}$ after correcting for the fiber-dependent instrumental resolution, we select the model including broad \ha.  Our choice of 20\% improvement in reduced $\chi^2$, while somewhat arbitrary, has been used successfully in previous studies \citep[e.g.,][]{haoetal2005} and empirically works here.  The modest FWHM requirement is imposed to avoid severe contamination from galaxies undergoing intense star formation that have moderately broadened bases on \ha.  The procedures described thus far yields a sample of 51 sources flagged as having a broad \ha\ component, {\it not} all of which make it into our final sample of broad-line AGN candidates as described in Section \ref{sec:broad}.

We also measure the \hbeta, \OIII\ $\lambda$5007 and \OI\ $6300$ emission lines.  The narrow-line profile derived from the \SII\ doublet is used as a template for fitting \hbeta, using the same approach as that for \ha.  The \hbeta\ line is fit twice, with and without a broad component.  If statistically justified (reduced $\chi^2$ is improved by at least 20\%), we accept the model with a broad component.  Since the \OIII\ profile commonly exhibits a broad, blue shoulder \citep[e.g.,][]{heckmanetal1981,whittle1985a} and does not typically match the profile of the other narrow lines measured in this work \citep{greeneho2005o3}, we do not use the model derived from the \SII\ doublet.  Instead, we fit the \OIII\ (and \OI) line independently allowing for up to two Gaussian components ($\chi^2$ must be reduced by at least 20\% to use the two-component model).  

Emission-line fluxes are calculated using the Gaussian model parameters.  Uncertainties in the model parameters are provided by MPFIT, which accounts for the SDSS error spectrum.  We use standard propagation of errors to determine the errors in the line fluxes.  

\subsection{BPT AGN and Composites}\label{sec:narrow}

\begin{figure*}[!t]
\begin{center}
{\includegraphics[width=6.9in]{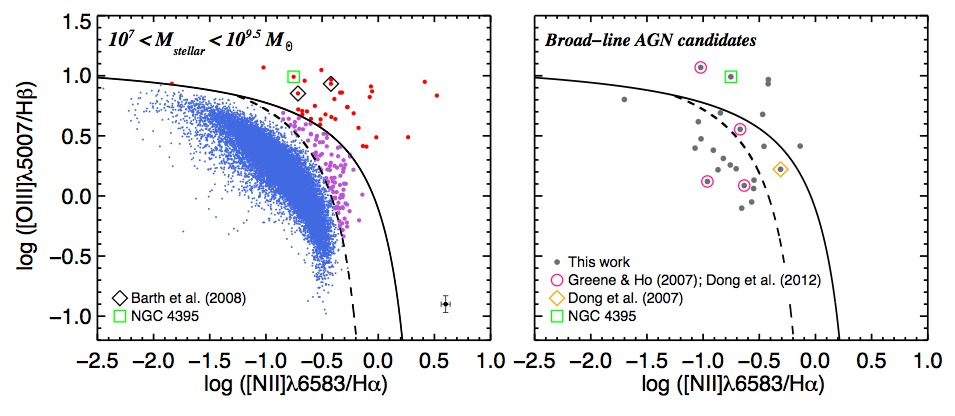}}
\end{center}
\caption{\footnotesize {\it Left:} BPT \OIII/\hbeta\ versus \NII/\ha\ narrow-line diagnostic diagram for all $\sim 25,000$ dwarf emission-line galaxies analyzed in this work.  Galaxies with spectra dominated by an AGN are plotted as red points and galaxies with spectra dominated by star formation are plotted as blue points.  Composite galaxies with significant contributions from both an AGN and star formation are plotted as purple points.  The typical error is shown in the lower right corner (individual flux errors are given in Tables \ref{tab:bptlines} and \ref{tab:sflines}).  The dashed line is an empirical separation of pure star-forming galaxies and those with some contribution from an AGN from \citet{kauffmannetal2003agn}.  The solid line is from \citet{kewleyetal2001}, indicating the `maximum starburst line' given by pure stellar photoionization models.  We note that the red point falling to the far left of the diagram and just above the dividing line is unusual in a number of ways and rather suspect (see footnote in text).      
{\it Right:} BPT diagram for broad-line AGN candidates only (\S\ref{sec:broad}).  We consider the sources falling in the AGN and composite regions of the diagram the most secure broad-line candidates.  Significant contamination from luminous Type II supernovae is likely in the star-forming region of the diagram.  
\label{fig:bpt}}
\end{figure*}

When an accreting BH is present in a galaxy, the ISM is photoionized by a much harder continuum than is emitted by hot stars.  AGNs and \ion{H}{2} regions separate cleanly in two-dimensional strong line diagnostic diagrams that take pairs of lines close together in frequency to mitigate the effects of reddening \citep{baldwinetal1981,veilleuxosterbrock1987,kewleyetal2001,kauffmannetal2003agn,kewleyetal2006}.  The two-dimensional line-diagnostic diagrams (also known as BPT diagrams after the Baldwin et al.\ paper) are routinely used to separate line-emitting galaxies by their primary excitation source.  

Here we employ the most widely used BPT diagram as our primary diagnostic, which takes \OIII/\hbeta\ vs.\ \NII/\ha\ (Figure \ref{fig:bpt}).  In this diagram, line-emitting galaxies separate into a $V$-shape \citep[e.g.,][]{kewleyetal2006, grovesetal2006}.  Star-forming galaxies with \ion{H}{2}-region-like spectra occupy the left-most plume of galaxies, with lower-metallicity systems having higher \OIII/\hbeta\ ratios and lower \NII/\ha\ ratios \citep[e.g.,][]{moustakasetal2006}.  AGNs occupy the right branch of galaxies with high-ionization Seyferts found in the upper right.  Low-metallicity AGNs are found to the left of the main Seyfert region and can overlap with low-metallicity starburst galaxies \citep{grovesetal2006,ludwigetal2012}.   The most contentious region of the diagram sits directly below the Seyfert galaxies: galaxies with very high low-ionization lines, named Low Ionization Nuclear Emission Region galaxies by \citet[LINERs; ][]{heckman1980}.  LINER emission can be generated both by shocks and very hard AGN spectra, and disentangling the primary origin in any given case can be complicated and aperture dependent \citep[e.g.,][]{kewleyetal2006,sarzietal2006,ho2008,eracleousetal2010,yanblanton2012}.  Finally, composites objects fall in the region delineated by the empirical dividing line of \citet{kauffmannetal2003agn} separating pure star-forming galaxies from those with some contribution from an AGN, and the theoretical `maximum starburst line' of \citet{kewleyetal2001} given by pure stellar photoionization models.  Composite objects are thought to contain significant contributions from both AGN and star formation \citep[e.g.,][but also see \citet{liuetal2008}]{panessaetal2005,kewleyetal2006,trouilleetal2011}.   

Using the \OIII/\hbeta\ vs.\ \NII/\ha\ diagnostic diagram and our narrow emission-line measurements, we have identified 136 galaxies with photoionization signatures of an active massive BH (Figure 1, Tables \ref{tab:bptgals} and \ref{tab:bptlines}).  There are 35 galaxies with narrow-line ratios falling in the AGN-dominated region of the BPT diagram and 101 galaxies with composite spectra.  While we consider the BPT-AGN more secure\footnote{The left-most red point in Figure \ref{fig:bpt} corresponds to a bright off-nuclear source in a blue late-type irregular dwarf galaxy, that may in fact be an extreme \HII\ region.  While we cannot exclude the possibility that this source is a low-metallicity active massive BH, we consider it rather suspect in the absence of other supporting evidence.}, the composite spectra likely indicate at least some contribution from an AGN \citep{trouilleetal2011,jiaetal2011}, which for our purposes is sufficient for tentatively identifying the presence of a massive BH.  Even with our inclusive approach, only $\sim 0.5\%~(136/25974)$ of dwarf emission-line galaxies in our parent sample exhibit optical narrow-line signatures of photoionization from an accreting BH (although see the discussion in Section \ref{sec:discussion} for a number of caveats).      

\begin{deluxetable*}{cccccccrcc}
\tabletypesize{\tiny}
\tablecaption{BPT AGN and Composites: Galaxy Properties}
\tablewidth{0pt}
\tablehead{
\colhead{ID} & \colhead{NSAID} & \colhead{SDSS Name} & \colhead{Plate-MJD-Fiber} &
\colhead{$z$}  & \colhead{log M$_\star$} & \colhead{$M_g$} & \colhead{$g-r$} & \colhead{$r_{50}$}  & \colhead{S{\'e}rsic $n$}  \\
\colhead{(1)} & \colhead{(2)} & \colhead{(3)} & \colhead{(4)} & \colhead{(5)} & \colhead{(6)} & \colhead{(7)} & \colhead{(8)} &
 \colhead{(9)} & \colhead{(10)} }
\startdata
\hline
\multicolumn{10}{c}{\it AGNs} \\
\hline
1 & 62996 & J024656.39$-$003304.8 & 1561-53032-161 & 0.0459 & 9.41 & $-17.99$ & $0.81$ & 1.52 & 1.0 \\
2 & 7480 & J024825.26$-$002541.4 & 409-51871-150 & 0.0247 & 9.11 & $-17.32$ & $0.58$ & 1.43 & 2.0 \\
3 & 68765 & J032224.64+401119.8 & 1666-52991-48 & 0.0261 & 9.42 & $-18.65$ & $0.47$ & 0.93 & 3.5 \\
4 & 64339 & J081145.29+232825.7 & 1584-52943-567 & 0.0157 & 9.02 & $-17.98$ & $0.36$ & 0.61 & 4.4 \\
5 & 46677 & J082334.84+031315.6 & 1185-52642-123 & 0.0098 & 8.54 & $-18.85$ & $-0.29$ & 1.66 & 2.8 \\
6 & 105376 & J084025.54+181858.9 & 2278-53711-445 & 0.0150 & 9.28 & $-17.61$ & $0.59$ & 1.00 & 3.8 \\
7 & 30020 & J084204.92+403934.5 & 829-52296-391 & 0.0293 & 9.30 & $-17.45$ & $0.62$ & 1.32 & 1.6 \\
8 & 110040 & J090222.76+141049.2 & 2433-53820-569 & 0.0297 & 9.41 & $-17.91$ & $0.59$ & 1.36 & 1.4 \\
9 & 10779 & J090613.75+561015.5 & 450-51908-409 & 0.0466 & 9.36 & $-18.98$ & $0.40$ & 1.63 & 4.1 \\
10 & 106134 & J092129.98+213139.3 & 2290-53727-185 & 0.0313 & 9.31 & $-18.20$ & $0.58$ & 1.07 & 6.0 \\
11 & 125318 & J095418.15+471725.1 & 2956-54525-457 & 0.0327 & 9.12 & $-18.73$ & $0.44$ & 2.00 & 6.0 \\
12 & 107272 & J100935.66+265648.9 & 2347-53757-456 & 0.0144 & 8.44 & $-17.56$ & $0.31$ & 0.66 & 6.0 \\
13 & 37942 & J102252.21+464220.7 & 961-52615-346 & 0.0400 & 9.44 & $-18.08$ & $0.66$ & 1.05 & 5.8 \\
14 & 111644 & J110503.96+224123.4 & 2488-54149-561 & 0.0246 & 9.05 & $-16.91$ & $0.74$ & 0.61 & 6.0 \\
15\tablenotemark{a} & 27397 & J110912.37+612347.0 & 774-52286-554 & 0.0068 & 8.91 & $-17.33$ & $0.36$ & 1.59 & 1.2 \\
16 & 30370 & J111319.23+044425.1 & 835-52326-433 & 0.0265 & 9.28 & $-17.98$ & $0.49$ & 1.31 & 4.3 \\
17 & 101949 & J114302.40+260818.9 & 2221-53792-218 & 0.0230 & 9.26 & $-17.39$ & $0.59$ & 0.74 & 1.9 \\
18 & 113566 & J114359.58+244251.7 & 2510-53877-255 & 0.0501 & 9.48 & $-18.68$ & $0.65$ & 1.32 & 5.6 \\
19 & 92369 & J114418.83+334007.4 & 2097-53491-228 & 0.0325 & 9.38 & $-17.76$ & $0.62$ & 0.87 & 6.0 \\
20\tablenotemark{b} & 52675 & J122342.82+581446.4 & 1315-52791-568 & 0.0144 & 9.47 & $-18.11$ & $0.66$ & 1.04 & 5.7 \\
21\tablenotemark{c} & 89394 & J122548.86+333248.7 & 2015-53819-251 & 0.0011 & 9.10 & $-18.05$ & $0.35$ & 1.71 & 1.7 \\
22 & 77431 & J130434.92+075505.0 & 1794-54504-547 & 0.0480 & 8.95 & $-18.77$ & $0.45$ & 1.23 & 5.6 \\
23 & 89494 & J130457.86+362622.2 & 2016-53799-414 & 0.0229 & 9.47 & $-18.64$ & $0.45$ & 3.94 & 0.8 \\
24 & 104527 & J133245.62+263449.3 & 2245-54208-467 & 0.0470 & 9.39 & $-19.07$ & $0.27$ & 1.18 & 2.3 \\
25 & 51220 & J134757.69+465434.9 & 1284-52736-95 & 0.0277 & 9.37 & $-18.13$ & $0.55$ & 0.62 & 6.0 \\
26 & 54572 & J134939.36+420241.4 & 1345-52814-41 & 0.0411 & 9.33 & $-18.55$ & $0.47$ & 0.89 & 3.0 \\
27 & 78568 & J140228.72+091856.4 & 1807-54175-560 & 0.0191 & 8.93 & $-16.57$ & $0.69$ & 0.87 & 1.6 \\
28 & 70907 & J140510.39+114616.9 & 1703-53799-510 & 0.0174 & 9.37 & $-18.52$ & $0.42$ & 1.30 & 4.5 \\
29 & 71023 & J141208.47+102953.8 & 1705-53848-98 & 0.0326 & 9.08 & $-17.74$ & $0.42$ & 0.62 & 3.9 \\
30 & 123656 & J142044.94+224236.8 & 2786-54540-552 & 0.0307 & 9.27 & $-18.35$ & $0.58$ & 1.07 & 5.9 \\
31 & 71565 & J143523.42+100704.2 & 1711-53535-306 & 0.0312 & 9.24 & $-18.53$ & $0.57$ & 0.97 & 2.9 \\
32\tablenotemark{a} & 15235 & J144012.70+024743.5 & 536-52024-575 & 0.0299 & 9.46 & $-19.18$ & $0.31$ & 1.70 & 2.7 \\
33 & 120870 & J144712.80+133939.2 & 2750-54242-275 & 0.0323 & 9.32 & $-18.31$ & $0.53$ & 1.54 & 1.8 \\
34 & 124249 & J153941.68+171421.8 & 2795-54563-509 & 0.0458 & 9.45 & $-18.99$ & $0.39$ & 1.19 & 3.6 \\
35 & 64022 & J154059.61+315507.3 & 1581-53149-255 & 0.0528 & 9.12 & $-19.36$ & $0.17$ & 1.33 & 6.0 \\
\hline
\multicolumn{10}{c}{\it Composites} \\
\hline
36 & 5790 & J002145.80+003327.3 & 390-51900-465 & 0.0178 & 9.16 & $-17.05$ & $0.65$ & 0.90 & 1.1 \\
37 & 6059 & J010005.93$-$011058.8 & 395-51783-58 & 0.0514 & 9.34 & $-18.76$ & $0.59$ & 1.19 & 5.7 \\
38 & 24013 & J024635.03+000719.1 & 707-52177-494 & 0.0286 & 9.06 & $-17.48$ & $0.50$ & 1.21 & 1.0 \\
39 & 24061 & J030644.60$-$002431.5 & 710-52203-279 & 0.0252 & 9.41 & $-18.16$ & $0.51$ & 1.18 & 0.9 \\
40 & 82616 & J074829.21+510052.4 & 1869-53327-263 & 0.0190 & 9.10 & $-17.89$ & $0.38$ & 3.05 & 1.7 \\
41 & 83933 & J080228.84+203050.3 & 1922-53315-364 & 0.0286 & 8.60 & $-19.85$ & $0.78$ & 2.04 & 2.0 \\
42 & 75073 & J081010.69+073337.1 & 1756-53080-413 & 0.0522 & 9.37 & $-19.53$ & $0.42$ & 1.88 & 6.0 \\
43 & 31779 & J081353.45+310824.3 & 861-52318-72 & 0.0479 & 9.41 & $-18.77$ & $0.72$ & 1.20 & 6.0 \\
44 & 50121 & J081549.08+254701.4 & 1266-52709-127 & 0.0248 & 9.18 & $-17.21$ & $0.63$ & 0.87 & 2.1 \\
45 & 35979 & J082013.92+302503.0 & 931-52619-145 & 0.0196 & 9.43 & $-17.97$ & $0.59$ & 1.12 & 1.7 \\
46 & 83280 & J083740.88+601339.3 & 1874-54452-158 & 0.0273 & 9.45 & $-18.72$ & $0.46$ & 2.00 & 0.5 \\
47 & 64526 & J084449.13+281853.5 & 1588-52965-176 & 0.0205 & 9.45 & $-17.84$ & $0.59$ & 0.93 & 2.5 \\
48 & 47066 & J085125.81+393541.7 & 1198-52669-392 & 0.0410 & 9.41 & $-19.19$ & $0.28$ & 1.76 & 0.5 \\
49 & 36154 & J085548.25+354502.0 & 935-52643-53 & 0.0527 & 9.45 & $-19.19$ & $0.42$ & 5.62 & 1.7 \\
50 & 47918 & J090737.05+352828.4 & 1212-52703-97 & 0.0276 & 9.44 & $-18.53$ & $0.53$ & 1.17 & 5.4 \\
\enddata
\tablecomments{Table \ref{tab:bptgals} is published in its entirety in the electronic edition of {\it The Astrophysical Journal}.
A portion is shown here for guidance regarding its form and content.  
Col.(1): Identification number assigned in this paper. Col.(2): NSA identification number. Col.(3): SDSS name. Col(4): Plate-MJD-Fiber of analyzed spectra.
Col.(5): Redshift.
Col.(6): Log galaxy stellar mass in units of M$_\odot$.
Col.(7): Absolute $g$-band magnitude.
Col.(8): $g-r$ color.
 Col.(9): Petrosian 50\% light radius in units of kpc. Col.(10): S{\'e}rsic index, $n$. All values are from the NSA and assume $h=0.73$.
 Magnitudes are $K$-corrected to rest-frame values using \texttt{kcorrect v4\_2} and corrected for foreground Galactic extinction.}
\tablenotetext{a}{Galaxies in Barth et al. (2008)}
\tablenotetext{b}{Galaxies in Greene \& Ho (2007) and Dong et al.\ (2012)}
\tablenotetext{c}{NGC 4395 \citep{filippenkosargent1989,filippenkoho2003}}
\tablenotetext{d}{Galaxy in Dong et al.\ (2007)}
\label{tab:bptgals}
\end{deluxetable*}

\begin{deluxetable*}{crrrrrrrrrr}
\hspace{-0.8cm}
\tabletypesize{\tiny}
\tablecaption{BPT AGN and Composites: Emission Line Fluxes}
\tablewidth{0pt}
\tablehead{
\colhead{ID} & \colhead{(H$\beta)_n$} & \colhead{(H$\beta)_b$} & \colhead{[O III]$\lambda$5007} & \colhead{[O I]$\lambda$6300} & \colhead{[N II]$\lambda$6548} & \colhead{(H$\alpha)_n$} & \colhead{(H$\alpha)_b$} & \colhead{[N II]$\lambda$6583} & \colhead{[S II]$\lambda$6716} & \colhead{[S II]$\lambda$6731} \\ 
\colhead{(1)} & \colhead{(2)} & \colhead{(3)} & \colhead{(4)} & \colhead{(5)} & \colhead{(6)} & \colhead{(7)} & \colhead{(8)} & \colhead{(9)} & \colhead{(10)} & \colhead{(11)} }
\startdata
\hline
\multicolumn{11}{c}{\it AGNs} \\
\hline
1 & 11(3) & \nodata & 29(3) & \nodata & 8(1) & 34(6) & 56(9) & 25(4) & 8(2) & 6(1) \\
2 & 13(3) & \nodata & 52(5) & \nodata & 6(1) & 61(3) & \nodata & 19(3) & 14(3) & 11(2) \\
3 & 100(11) & \nodata & 1119(61) & 45(3) & 44(5) & 418(45) & \nodata & 130(6) & 64(8) & 67(11) \\
4 & 82(6) & \nodata & 460(75) & 57(6) & 28(2) & 299(7) & \nodata & 82(5) & 74(5) & 68(4) \\
5 & 5813(69) & \nodata & 49827(429) & 144(6) & 100(2) & 20342(193) & \nodata & 296(5) & 520(9) & 420(6) \\
6 & 101(6) & \nodata & 515(11) & 20(4) & 22(1) & 313(8) & \nodata & 66(4) & 71(5) & 50(3) \\
7 & 13(2) & \nodata & 62(4) & \nodata & 13(1) & 62(5) & \nodata & 38(4) & 25(3) & 20(2) \\
8 & 19(3) & \nodata & 58(15) & 13(4) & 29(2) & 90(7) & \nodata & 87(7) & 43(5) & 31(3) \\
9 & 151(13) & 79(12) & 1398(34) & 121(10) & 68(5) & 530(42) & 322(14) & 202(10) & 97(9) & 79(13) \\
10 & 34(3) & \nodata & 155(6) & \nodata & 17(1) & 131(5) & \nodata & 50(3) & 47(4) & 30(2) \\
11 & 510(20) & \nodata & 2433(44) & 179(10) & 188(7) & 1643(66) & 118(16) & 557(18) & 267(16) & 251(26) \\
12 & 336(8) & \nodata & 3048(87) & 226(6) & 90(2) & 1199(16) & \nodata & 265(5) & 233(5) & 182(4) \\
13 & 22(4) & \nodata & 56(6) & \nodata & 20(2) & 77(7) & \nodata & 60(6) & 45(6) & 33(4) \\
14 & 15(3) & \nodata & 109(6) & \nodata & 10(1) & 63(5) & \nodata & 30(4) & 21(4) & 13(2) \\
15 & 93(5) & \nodata & 660(11) & 40(3) & 21(1) & 327(6) & \nodata & 63(2) & 87(3) & 72(2) \\
16 & 33(4) & \nodata & 171(6) & \nodata & 8(1) & 128(4) & \nodata & 25(2) & 34(3) & 22(2) \\
17 & 125(9) & \nodata & 459(13) & 47(6) & 121(6) & 535(12) & \nodata & 357(17) & 84(6) & 83(4) \\
18 & 22(4) & \nodata & 120(23) & \nodata & 17(2) & 94(5) & \nodata & 50(6) & 23(4) & 16(2) \\
19 & 23(6) & \nodata & 154(18) & 27(5) & 33(3) & 117(11) & \nodata & 98(9) & 43(6) & 32(4) \\
20 & 211(13) & 495(37) & 2477(49) & 120(10) & 52(4) & 1604(82) & 3195(50) & 153(11) & 98(7) & 101(6) \\
21 & 3314(47) & 893(39) & 32481(394) & 3542(49) & 777(11) & 13026(162) & 6167(73) & 2300(30) & 2434(35) & 2994(59) \\
22 & 16(3) & \nodata & 140(6) & \nodata & 42(4) & 48(4) & \nodata & 125(12) & 29(4) & 18(2) \\
23 & 39(5) & \nodata & 215(7) & 18(3) & 23(1) & 128(4) & \nodata & 68(4) & 52(4) & 37(2) \\
24 & 539(22) & \nodata & 3905(67) & 238(15) & 317(13) & 2048(83) & \nodata & 937(27) & 394(18) & 367(22) \\
25 & 151(9) & \nodata & 1034(183) & 78(8) & 671(21) & 597(13) & \nodata & 1985(63) & 165(8) & 124(5) \\
26 & 93(8) & \nodata & 691(16) & 26(7) & 102(5) & 344(10) & \nodata & 303(16) & 74(6) & 61(4) \\
27 & 32(4) & \nodata & 202(6) & 26(3) & 22(1) & 158(4) & \nodata & 64(3) & 53(3) & 41(2) \\
28 & 81(7) & \nodata & 353(11) & 56(7) & 42(2) & 299(9) & \nodata & 124(7) & 102(7) & 73(4) \\
29 & 63(6) & \nodata & 516(12) & 55(5) & 82(3) & 280(7) & \nodata & 241(10) & 82(5) & 64(4) \\
30 & 33(5) & \nodata & 142(6) & \nodata & 9(1) & 87(4) & \nodata & 27(4) & 21(4) & 21(3) \\
31 & 156(8) & \nodata & 682(33) & 30(8) & 49(3) & 581(12) & \nodata & 146(10) & 62(6) & 51(4) \\
32 & 832(19) & 132(15) & 7136(94) & 375(13) & 448(8) & 3496(66) & 301(36) & 1326(19) & 502(11) & 495(15) \\
33 & 23(3) & \nodata & 72(6) & 18(6) & 49(4) & 79(7) & \nodata & 145(12) & 53(5) & 42(4) \\
34 & 180(6) & \nodata & 906(38) & 15(4) & 47(2) & 597(10) & \nodata & 138(5) & 90(4) & 66(3) \\
35 & 470(20) & \nodata & 2233(90) & 90(5) & 118(5) & 1661(64) & \nodata & 351(8) & 310(13) & 228(12) \\
\hline
\multicolumn{11}{c}{\it Composites} \\
\hline
36 & 52(4) & \nodata & 115(4) & 27(3) & 40(1) & 212(4) & \nodata & 119(4) & 67(3) & 49(2) \\
37 & 299(23) & \nodata & 878(101) & 64(5) & 105(8) & 1113(83) & \nodata & 312(16) & 216(26) & 160(33) \\
38 & 14(3) & \nodata & 18(4) & \nodata & 5(1) & 43(6) & \nodata & 15(2) & 15(3) & 11(2) \\
39 & 35(4) & \nodata & 40(5) & \nodata & 17(1) & 134(5) & \nodata & 51(4) & 34(3) & 28(3) \\
40 & 19(3) & \nodata & 86(5) & 6(2) & 5(1) & 82(4) & \nodata & 16(2) & 20(3) & 13(2) \\
41 & 83(8) & \nodata & 71(10) & \nodata & 65(5) & 362(12) & \nodata & 191(15) & 88(9) & 53(6) \\
42 & 327(44) & \nodata & 618(47) & 18(5) & 161(22) & 1260(168) & \nodata & 477(45) & 102(18) & 98(30) \\
43 & 41(5) & \nodata & 36(5) & \nodata & 20(2) & 133(6) & \nodata & 59(5) & 36(4) & 22(3) \\
44 & 109(14) & \nodata & 118(9) & 38(6) & 66(8) & 495(59) & \nodata & 197(35) & 140(26) & 106(34) \\
45 & 18(5) & \nodata & 27(6) & \nodata & 9(1) & 61(5) & \nodata & 28(4) & 22(4) & 25(4) \\
46 & 24(4) & \nodata & 13(4) & \nodata & 15(1) & 97(4) & \nodata & 44(4) & 26(3) & 15(2) \\
47 & 27(4) & \nodata & 32(5) & 16(4) & 18(1) & 133(5) & \nodata & 53(4) & 38(4) & 22(3) \\
48 & 169(7) & \nodata & 438(10) & 31(5) & 61(2) & 518(11) & 137(18) & 180(6) & 110(5) & 75(3) \\
49 & 8(3) & \nodata & 18(4) & 0() & 4(1) & 48(3) & \nodata & 13(2) & 16(2) & 10(1) \\
50 & 107(7) & \nodata & 350(10) & 25(6) & 42(2) & 358(9) & \nodata & 125(6) & 93(6) & 73(4) \\
\enddata
\tablecomments{Table \ref{tab:bptlines} is published in its entirety in the electronic edition of {\it The Astrophysical Journal}.
A portion is shown here for guidance regarding its form and content. 
Col.(1): Identification number assigned in this paper. Col.(2)-(11): Emission line fluxes 
with units of $10^{-17}$ erg s$^{-1}$ cm$^{-2}$.  Errors are shown in parenthesis.
No extinction correction has been applied.
The subscripts $n$ and $b$ refer to the narrow and broad components of the line, respectively. A three-dot ellipsis indicates that no line is detected. In the cases where a broad component of \ha\ or \hbeta\ is detected, we use the flux of the narrow component to calculate the emission line ratios for the BPT diagram.}
\label{tab:bptlines}
\end{deluxetable*}

\begin{figure}[!t]
\begin{center}
{\includegraphics[width=3.1in]{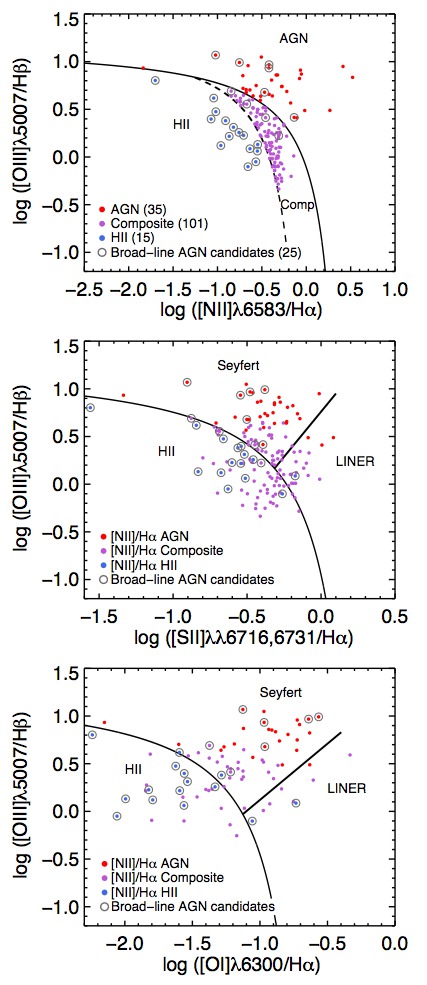}}
\end{center}
\caption{\small BPT narrow-line diagnostic diagrams for our sample of active galaxies.  Regions are delineated according to the classification scheme outlined in \citet{kewleyetal2006}.  {\it Top:} \OIII/\hbeta\ versus \NII/\ha\ diagram.  There are 35 galaxies in the AGN part of the diagram (6 with broad \ha\ emission) and 101 galaxies in the composite region of the diagram (4 with broad \ha\ emission).  An additional 15 galaxies have broad \ha\ emission, yet have HII-region-like narrow-line ratios.
{\it Middle:} \OIII/\hbeta\ versus \SII/\ha\ diagram.  Colors indicate classification based on the OIII/\hbeta\ versus \NII/\ha\ diagram.
{\it Bottom:} \OIII/\hbeta\ versus \OI/\ha\ diagram for galaxies in which we detect the \OI\ emission line (Tables \ref{tab:bptlines} and \ref{tab:sflines}).  Colors indicate classification based on the OIII/\hbeta\ versus \NII/\ha\ diagram.
\label{fig:allbpt}}
\end{figure}

We also investigate the positions of AGNs and composite galaxies according to the \OIII/\hbeta\ vs.\ \NII/\ha\ diagram in two secondary diagnostic diagrams, \OIII/\hbeta\ vs.\ \SII/\ha\ and \OIII/\hbeta\ vs.\ \OI/\ha\ (Figure \ref{fig:allbpt}).  \OI/\ha\ is particularly useful as it is sensitive to the hardness of the ionizing radiation field \citep[e.g.,][]{kewleyetal2006}, although \OI\ is relatively weak and not detected in $\sim$50\% of the AGNs + composites (Table \ref{tab:bptlines}).  Figure \ref{fig:allbpt} shows that nearly all of the \NII/\ha\ AGN fall in the Seyfert region of the secondary diagnostic diagrams, with at most 3 falling in the LINER part of the diagrams.  The \NII/\ha\ composites fall throughout the \SII/\ha\ and \OI/\ha\ diagrams.

\begin{figure*}[!h]
\epsscale{1.1}
\hspace{-0.1cm}
\plotone{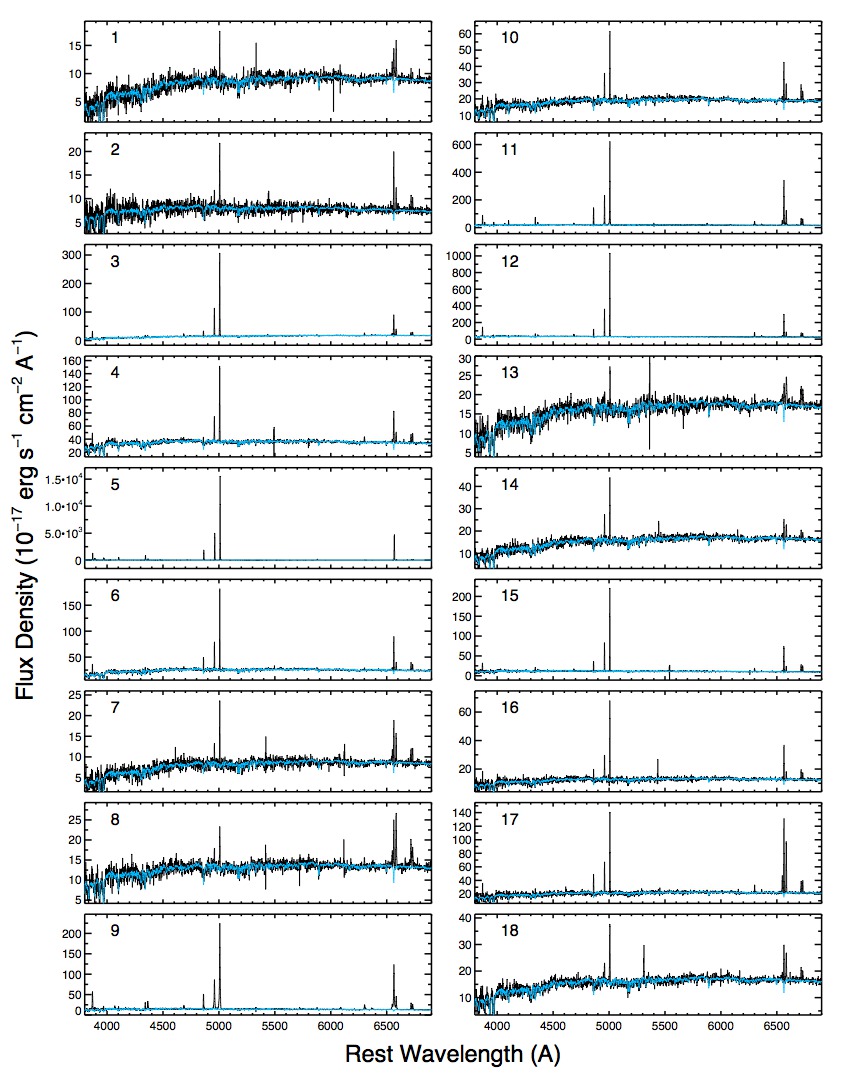}
\caption{\footnotesize SDSS redshift-corrected spectra of galaxies falling in the AGN region of the \OIII/\hbeta\ versus \NII/\ha\ diagram.  Continuum and absorption-line fits are shown in
blue (see Section \ref{sec:contsub}).  An identification number (Table \ref{tab:bptgals}) is given in the upper left corner of each plot.}
\label{fig:nlagn_spec1}
\end{figure*}

\begin{figure*}[!t]
\epsscale{1.1}
\hspace{-0.1cm}
\plotone{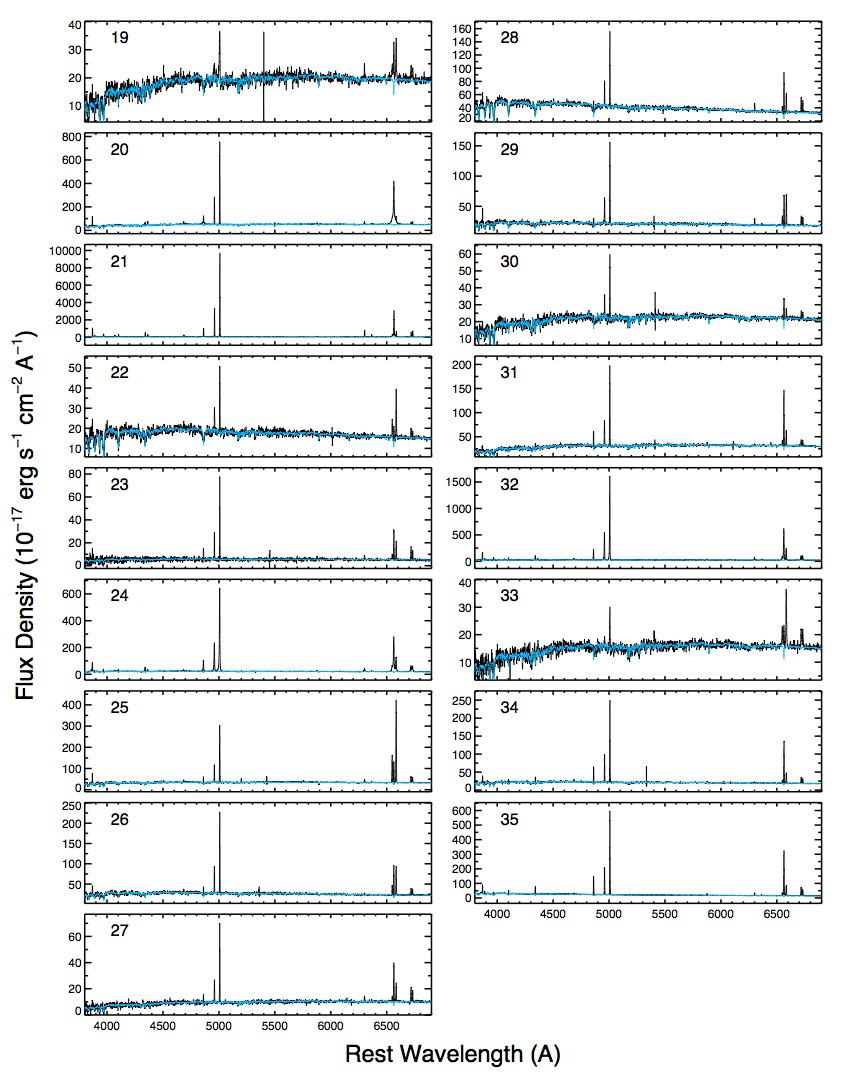}
\caption{\footnotesize Same as Figure \ref{fig:nlagn_spec1}.}
\label{fig:nlagn_spec2}
\end{figure*}

The SDSS spectra of the BPT-AGN are shown in Figures \ref{fig:nlagn_spec1} and \ref{fig:nlagn_spec2} with the continuum and absorption line models over-plotted in blue.  Our sample of BPT-AGN includes NGC 4395 and 2 galaxies from the low-mass Seyfert 2 sample of \citet{barthetal2008a}.  The remaining galaxies in \citet{barthetal2008a} do not meet the selection criteria to be included in our parent dwarf galaxy sample (\S\ref{sec:sample}).  Due to the large number of composites, we do not show their spectra here.  However, we have inspected all of the individual spectra flagged as composites by eye and cut sources with unreliable emission line measurements due to low signal-to-noise.  Twenty objects were excluded, leaving a final sample of 101 composites.

\subsection{Broad-Line AGN Candidates}\label{sec:broad}

We also search for broad \ha\ emission in our parent sample of galaxies, which may indicate dense gas orbiting a BH within the broad-line region (BLR), only light-days from the central BH \citep[e.g.,][]{petersonetal2004,petersonetal2005,bentzetal2009lamp}.  Unobscured quasars powered by $\sim 10^8$~\msun\ BHs have typical line-widths of $\sim$3000 km s$^{-1}$ \citep[e.g.,][]{shenetal2008b}.  However, in AGNs with low-mass BHs ($M_{\rm BH} \lesssim 10^{6.5}$~\msun), the line-widths can be just hundreds of km s$^{-1}$ \citep[e.g.,][]{filippenkoho2003, greeneho2007}.  While in principle broad emission lines provide clear evidence that gas is moving in the potential of a compact massive object \citep[e.g.,][]{hfs1997broad}, there are a few complications, particularly in this low-mass regime.  First, \ha\ absorption from young stars can mask or mimic the broad \ha, and thus accurate galaxy continuum subtraction is required \citep[Section \ref{sec:contsub}; also see][]{greeneho2004}. Second, some varieties of supernovae (SNe) can masquerade as AGNs at low-luminosities (see below).
 
\begin{figure*}[!t]
\epsscale{1.15}
\plotone{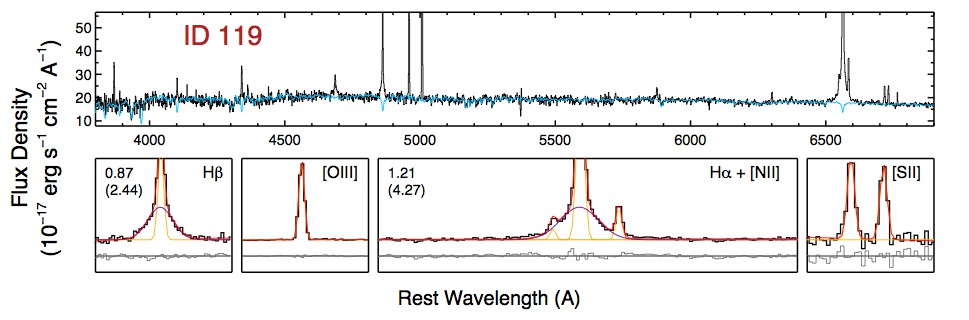}
\caption{\footnotesize Example of a broad-line AGN candidate (ID 119).  The narrow-line ratios of this source place it in the composite section of the \OIII/\hbeta\ versus \NII/\ha\ diagram.  {\it Top:} The redshift-corrected spectrum with the continuum and absorption-line fit plotted in blue.  {\it Bottom:} Chunks of the emission-line spectrum (after continuum and absorption-line subtraction).  Best-fitting models are plotted
in red and the individual narrow Gaussian components are plotted in yellow.  The broad \ha\ and \hbeta\ components are plotted in dark blue.  Residuals are plotted in gray with a vertical offset for clarity.  Reduced $\chi^2$ values are shown in the upper left-hand corner of the \ha\ + [N II] and \hbeta\ chunks.  For comparison, the reduced $\chi^2$ values from the fits not including a broad component are shown in parenthesis.  Spectra of the other 24 broad-line AGN candidates are shown in the Appendix.}
\label{fig:broadex}
\end{figure*}

Out of the 51 sources flagged as having a broad \ha\ component in their spectra with a FWHM $\geq$ 500 km s$^{-1}$ (see \S\ref{sec:linefit}), 25 make it into our final sample of broad-line AGN candidates after a more careful examination of each individual source.  We detect broad \hbeta, in addition to broad \ha\ in $36\%$ (9/25) of these sources (Tables \ref{tab:bptlines} and \ref{tab:sflines}).  We have excluded 9 likely Type II SNe (most having characteristic P Cygni profiles), 3 non-nuclear star-forming knots in nearby galaxies (with low broad \ha\ luminosities), 1 Luminous Blue Variable star identified by \citet[][NSAID 5109]{izotovthuan2009lbv}, and 13 sources in the star-forming region of the BPT diagram with marginal detections of broad \ha\ that are unconvincing by eye.  Figure \ref{fig:broadex} shows the spectral fits for one of our broad-line AGN candidates.  Plots for the other broad-line AGN candidates are shown in the Appendix.

Our broad-line sample includes NGC 4395, the dwarf disk galaxy presented in \citet{dongetal2007}, and 4 galaxies from the samples of \citet{greeneho2007} and \citet{dongetal2012}.  The remaining samples of low-mass BHs in the latter two works are hosted in galaxies that do not meet the selection criteria to be included in our parent dwarf catalog (\S\ref{sec:sample}).  The same is true of the four candidate metal-poor AGN presented in \citet{izotovthuan2008}.  Our broad-line sample also includes HS 0837$+$4717 (source B), which has narrow-line ratios consistent with a low-metallicity starburst or a low-metallicity AGN, and exhibits persistent broad emission lines \citep{kniazevetal2000,izotovetal2007}.  

\begin{figure}[!h]
\epsscale{1.25}
\hspace{-0.8cm}
\plotone{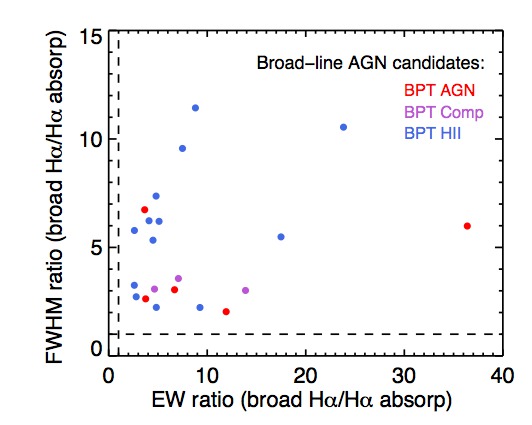}
\caption{\footnotesize Ratios of the EWs and FWHMs of the broad components of \ha\ emission to the \ha\ absorption lines for the broad-line AGN candidates (Section \ref{sec:broad}).  Points are color-coded according to their position in the \OIII/\hbeta\ versus \NII/\ha\ BPT diagram.  Dashed lines indicate ratios equal to 1.  This plot demonstrates that the broad \ha\ emission features are not a result of over-subtracting the \ha\ absorption features.   Absorption lines were not detected in 4 of the 25 broad-line AGN candidates and they are excluded here. }\label{fig:ew}
\end{figure}

For each of our broad-line candidates, we check if the broad \ha\ component could be an artifact from over-subtracting the \ha\ absorption line.  We fit a single Gaussian to the \ha\ absorption line in the model fit (Section \ref{sec:contsub}), measure the corresponding equivalent width (EW) and FWHM, and compare these values to the broad \ha\ emission component.  As shown in Figure \ref{fig:ew}, the EWs of the broad emission features are $\sim$ 3 to 36 times the EWs of the absorption features and the FWHMs of the broad \ha\ components are $\sim$ 2 to 11 times the FWHMs of the absorption lines.  Moreover, in all but 1 case, the FWHMs of the \ha\ absorption lines are less than 500 km s$^{-1}$, our minimum threshold for the width of any broad \ha\ emission.  Thus, our broad \ha\ detections appear to be robust and not a result of over-subtracting an absorption feature.

The broad-line AGN candidates are found throughout the BPT diagrams (Figures \ref{fig:bpt} and \ref{fig:allbpt}), as is the case for other samples of Type 1 AGN \citep{greeneho2007,sternlaor2013}.  Ten of the galaxies fall in either the AGN or composite region of the \OIII/\hbeta\ vs.\ \NII/\ha\ diagram and we consider these the most secure broad-line AGN candidates, half of which are new identifications\footnote{\citet{barthetal2008a} identify a tentative broad component in ID 32 (Table \ref{tab:bptgals}).}.  Accounting for just the BPT-AGN, the fraction of sources with a detectable broad \ha\ component (i.e., the Type 1 fraction) is $\sim$17\%~(6/35).  The other 15 broad-line sources fall in the star-forming region of the \OIII/\hbeta\ vs.\ \NII/\ha\ diagram (Tables \ref{tab:sfgals} and \ref{tab:sflines}), 2 of which fall in the Seyfert (and 2 in the LINER) region of the \OIII/\hbeta\ vs.\ \OI/\ha\ diagram.  While models of low-metallicity AGNs overlap with low-metallicity starbursts \citep{grovesetal2006}, only one of the broad-line sources in the star-forming part of the \OIII/\hbeta\ vs.\ \NII/\ha\ diagram also has narrow-line ratios consistent with a low-metallicity AGN (source B).  For the majority of cases, the narrow-lines are likely dominated by star formation.  Bona fide broad-line AGN falling in the star-forming part of the diagnostic diagram can naturally be explained by star-formation dominating the narrow-line emission within the 3\arcsec\ SDSS aperture, which can cover a substantial fraction of the host galaxy for these dwarfs.  It is also possible, however, that the broad \ha\ seen in objects lying in the star-forming region of the BPT diagram is in fact from stellar phenomena.  

\begin{deluxetable*}{cccccccrcc}
\tabletypesize{\tiny}
\tablecaption{BPT Star-Forming Galaxies with Broad H$\alpha$: Galaxy Properties}
\tablewidth{0pt}
\tablehead{
\colhead{ID} & \colhead{NSAID} & \colhead{SDSS Name} & \colhead{Plate-MJD-Fiber} &
\colhead{$z$}  & \colhead{log M$_\star$} & \colhead{$M_g$} & \colhead{$g-r$} & \colhead{$r_{50}$}  & \colhead{S{\'e}rsic $n$}  \\
\colhead{(1)} & \colhead{(2)} & \colhead{(3)} & \colhead{(4)} & \colhead{(5)} & \colhead{(6)} & \colhead{(7)} & \colhead{(8)} &
 \colhead{(9)} & \colhead{(10)} }
\startdata
A\tablenotemark{a} & 22083 & J004042.10$-$110957.7 & 655-52162-89 & 0.0274 & 9.45 & $-18.14$ & $0.56$ & 1.0 & 2.3 \\
B\tablenotemark{b} & 15952 & J084029.91+470710.4 & 549-51981-621 & 0.0421 & 8.11 & $-18.89$ & $-0.85$ & 0.9 & 6.0 \\
C & 109990 & J090019.66+171736.9 & 2432-54052-524 & 0.0288 & 9.30 & $-19.08$ & $0.26$ & 2.3 & 1.1 \\
D & 76788 & J091122.24+615245.5 & 1786-54450-514 & 0.0266 & 8.79 & $-18.56$ & $0.26$ & 2.2 & 0.6 \\
E & 109016 & J101440.21+192448.9 & 2373-53768-148 & 0.0289 & 8.75 & $-17.90$ & $0.23$ & 0.9 & 1.9 \\
F & 12793 & J105100.64+655940.7 & 490-51929-279 & 0.0325 & 9.11 & $-19.05$ & $0.11$ & 0.8 & 6.0 \\
G & 13496 & J105447.88+025652.4 & 507-52353-619 & 0.0222 & 8.90 & $-18.56$ & $0.20$ & 0.9 & 2.9 \\
H & 74914 & J111548.27+150017.7 & 1752-53379-532 & 0.0501 & 8.82 & $-18.93$ & $0.19$ & 1.5 & 5.3 \\
I & 112250 & J112315.75+240205.1 & 2497-54154-221 & 0.0250 & 9.01 & $-18.33$ & $0.44$ & 0.8 & 6.0 \\
J\tablenotemark{a} & 41331 & J114343.77+550019.4 & 1015-52734-596 & 0.0272 & 9.01 & $-17.97$ & $0.22$ & 1.1 & 0.9 \\
K & 91579 & J120325.66+330846.1 & 2089-53498-283 & 0.0349 & 9.01 & $-17.44$ & $0.92$ & 1.1 & 5.9 \\
L & 33207 & J130724.64+523715.5 & 887-52376-454 & 0.0262 & 9.09 & $-19.14$ & $0.19$ & 1.2 & 1.3 \\
M & 119311 & J131503.77+223522.7 & 2651-54507-488 & 0.0230 & 9.14 & $-18.90$ & $0.29$ & 1.7 & 3.9 \\
N & 88972 & J131603.91+292254.0 & 2009-53904-640 & 0.0378 & 8.93 & $-19.72$ & $-0.14$ & 1.4 & 3.1 \\
O & 104565 & J134332.09+253157.7 & 2246-53767-49 & 0.0287 & 9.35 & $-18.58$ & $0.37$ & 3.3 & 0.9 \\
\enddata
\tablecomments{Col.(1): Identification assigned in this paper. Col.(2): NSA identification number. 
Col.(3): SDSS name. Col.(4): Plate-MJD-Fiber of analyzed spectra.
Col.(5): Redshift.
Col.(6): Log galaxy stellar mass in units of M$_\odot$.
Col.(7): Absolute $g$-band magnitude.
Col.(8): $g-r$ color.
 Col.(9): Petrosian 50\% light radius in units of kpc. Col.(10): S{\'e}rsic index, $n$.
All values are from the NSA and assume $h=0.73$.
Magnitudes are $K$-corrected to rest-frame values using \texttt{kcorrect v4\_2} and corrected for foreground Galactic extinction.}
\tablenotetext{a}{Galaxies in Greene \& Ho (2007) and Dong et al.\ (2012)}
\tablenotetext{b}{HS 0837$+$4717 \citep{izotovetal2007}.  The [O III]/\hbeta\ and [N II]/\ha\ ratios for this source are also consistent with a low-metallicity AGN.}
\label{tab:sfgals}
\end{deluxetable*}

\begin{deluxetable*}{crrrrrrrrrr}
\hspace{-0.8cm}
\tabletypesize{\tiny}
\tablecaption{BPT Star-Forming Galaxies with Broad H$\alpha$: Emission Line Fluxes}
\tablewidth{0pt}
\tablehead{
\colhead{ID} & \colhead{(H$\beta)_n$} & \colhead{(H$\beta)_b$} & \colhead{[O III]$\lambda$5007} & \colhead{[O I]$\lambda$6300} & \colhead{[N II]$\lambda$6548} & \colhead{(H$\alpha)_n$} & \colhead{(H$\alpha)_b$} & \colhead{[N II]$\lambda$6583} & \colhead{[S II]$\lambda$6716} & \colhead{[S II]$\lambda$6731} \\ 
\colhead{(1)} & \colhead{(2)} & \colhead{(3)} & \colhead{(4)} & \colhead{(5)} & \colhead{(6)} & \colhead{(7)} & \colhead{(8)} & \colhead{(9)} & \colhead{(10)} & \colhead{(11)} }
\startdata
A & 42(5) & \nodata & 52(6) & 21(5) & 9(1) & 113(6) & 218(17) & 26(3) & 44(4) & 32(3) \\
B & 2287(47) & 195(20) & 14521(202) & 46(4) & 54(2) & 8066(82) & 1292(23) & 161(5) & 115(4) & 107(3) \\
C & 340(6) & \nodata & 560(9) & 29(3) & 53(1) & 1155(14) & 749(21) & 157(3) & 190(4) & 141(3) \\
D & 180(6) & \nodata & 450(10) & 16(4) & 16(1) & 571(9) & 251(26) & 48(3) & 99(4) & 66(3) \\
E & 408(10) & \nodata & 1218(24) & 32(5) & 44(1) & 1348(20) & 107(17) & 129(4) & 170(6) & 125(4) \\
F & 917(79) & 311(13) & 1241(39) & 38(5) & 361(31) & 3764(323) & 582(28) & 1069(77) & 293(57) & 263(89) \\
G & 1058(19) & \nodata & 1782(167) & 54(6) & 237(4) & 3567(45) & 365(27) & 702(12) & 518(11) & 373(7) \\
H & 164(6) & \nodata & 394(10) & 28(5) & 23(1) & 549(10) & 183(22) & 68(4) & 86(4) & 64(3) \\
I & 576(14) & \nodata & 664(15) & 62(7) & 215(4) & 2264(30) & 196(27) & 638(12) & 392(10) & 305(7) \\
J & 227(9) & 80(18) & 299(10) & 17(5) & 39(2) & 1065(31) & 648(22) & 116(5) & 135(6) & 90(4) \\
K & 183(5) & \nodata & 330(6) & 30(3) & 39(1) & 655(9) & 155(10) & 114(3) & 133(4) & 95(2) \\
L & 685(16) & \nodata & 1405(26) & 63(7) & 111(3) & 2174(31) & 236(35) & 329(8) & 371(10) & 288(7) \\
M & 495(11) & \nodata & 441(20) & 15(5) & 154(4) & 1690(22) & 521(30) & 457(10) & 229(7) & 169(5) \\
N & 3524(44) & \nodata & 14615(162) & 302(16) & 365(4) & 11905(129) & 907(55) & 1080(13) & 981(15) & 727(10) \\
O & 34(4) & \nodata & 27(4) & 9(2) & 8(1) & 106(3) & 46(8) & 23(2) & 36(3) & 22(2) \\
\enddata
\tablecomments{Col.(1): Identification assigned in this paper. Col.(2)-(11): Emission line fluxes with units of $10^{-17}$ erg s$^{-1}$ cm$^{-2}$.
Errors are shown in parenthesis.  No extinction correction has been applied.
The subscripts $n$ and $b$ refer to the narrow and broad components of the line, respectively. A three-dot ellipsis indicates that no broad component of H$\beta$ is detected.}
\label{tab:sflines}
\end{deluxetable*}

\begin{figure}[!h]
\epsscale{1.1}
\hspace{-0.8cm}
\plotone{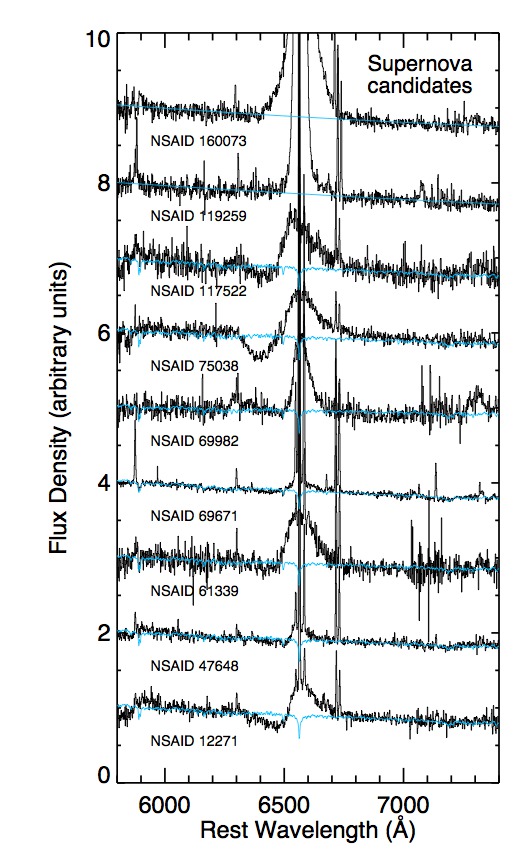}
\caption{\footnotesize The 9 supernovae candidates excluded from our broad-line AGN candidates.  The spectral region around \ha\ is shown with the continuum and absorption line fits plotted in blue. }\label{fig:sne}
\end{figure}

Broad \ha\ from galaxies in the star-forming region of the BPT diagram may well be from luminous Type II SNe that happened to be detectable when the SDSS spectra were taken and fell within the spectroscopic aperture.  Type II SNe can exhibit broad \ha\ emission with luminosities upwards of $\sim 10^{40}$ erg s$^{-1}$, which is comparable to the luminosities of our broad-line sources and other examples of AGNs with low-mass BHs \citep{filippenkoho2003,greeneho2007}.  Some Type II SNe also exhibit broad P Cygni profiles in \ha\ and we have already excluded these sources from our broad-line sample.  We identified 9 such objects by eye (Table \ref{tab:sn}), and these were subsequently confirmed by the automated SNe detection code used in \citet{graurmaoz2013} (O.Graur, private communication).  Two objects, NSAID 119259 and NSAID 69982,  have also been identified as Type II SNe by \citet{izotovthuan2009sn} and \citet{izotovetal2007}, respectively.  Figure \ref{fig:sne} shows the spectral region around \ha\ for the SNe candidates.  In some cases, the P Cygni profile is very subtle with only slight blue-shifted absorption indicated by an asymmetric emission line profile.  Another way we can identify SNe in our broad-line sample is to examine the temporal evolution of the broad \ha\ emission.  We would expect the broad line to persist for an AGN, whereas it should significantly decrease or disappear over a timespan of several years for SNe.  Therefore, we are currently obtaining follow-up spectroscopy of the broad-line sources, the results of which will be presented in a forthcoming paper. 

\begin{deluxetable*}{cccccc}
\tabletypesize{\tiny}
\tablecaption{Supernova Candidates}
\tablewidth{0pt}
\tablehead{
\colhead{NSAID} & \colhead{SDSS Name} & \colhead{Plate-MJD-Fiber} &
\colhead{$z$} & \colhead{log M$_\star$} & \colhead{$M_g$} \\
\colhead{(1)} & \colhead{(2)} & \colhead{(3)} & \colhead{(4)} & \colhead{(5)} & \colhead{(6)} }
\startdata
12271 & J093313.94+015858.7 & 475-51965-626 & 0.0311 & 8.94 & $-$17.81 \\
47648 & J082449.94+293644.1 & 1207-52672-512 & 0.0404 & 9.45 & $-$19.06 \\
61339 & J131307.11+460554.3 & 1459-53117-22 & 0.0296 & 9.46 & $-$18.40 \\
69671 & J162244.78+323933.0 & 1684-53239-484 & 0.0410 & 9.41 & $-$19.74 \\
69982\tablenotemark{a} & J164402.63+273405.4 & 1690-53475-360 & 0.0232 & 8.93 & $-$17.80 \\
75038 & J113913.54+150215.7 & 1755-53386-516 & 0.0140 & 8.51 & $-$17.08 \\
117522 & J103134.64+190407.1 & 2593-54175-334 & 0.0342 & 8.89 & $-$17.92 \\
119259\tablenotemark{b} & J132053.66+215510.2 & 2651-54507-31 & 0.0224 & 8.55 & $-$16.97 \\
160073 & J113322.89+550420.0 & 1014-52707-463 & 0.0091 & 8.65 & $-$18.16 \\
\enddata
\tablecomments{Col.(1): NSA identification number. Col.(2): SDSS name. Col(3): Plate-MJD-Fiber of analyzed spectra.
Col.(4): Redshift from NSA. Col.(5): Galaxy stellar mass from NSA corrected for $h = 0.73$. Col.(6): Absolute $g$-band magnitude
from NSA corrected for $h = 0.73$ and foreground Galactic extinction.}
\tablenotetext{a}{Possible Type IIp SNe identified by Izotov et al.\ (2007)}
\tablenotetext{b}{Type IIn SNe identified by Izotov \& Thuan (2009)}
\label{tab:sn}
\end{deluxetable*}

We also consider stellar winds from evolving massive stars undergoing mass loss, such as those from Wolf-Rayet (WR) stars and Luminous Blue Variables (LBVs).  WR stars are identified in the integrated spectra of galaxies by the blended emission from helium, carbon and nitrogen at $\lambda 4650-4690$ known as the ``WR bump" and this feature is only detected in four of our broad-line AGN candidates by \citet{brinchmannetal2008}\footnote{The work of \citet{brinchmannetal2008} makes use of DR6, whereas we use DR8.  18 of our 25 broad-line AGN candidates are found in DR6.}.  Two are BPT-AGN, including NGC 4395 and one of the galaxies from the \citet{barthetal2008a} sample (IDs 21 and 32), and two are BPT-star-forming galaxies (IDs B and F).  We note that broad \HeII\ $\lambda 4686$ can also be produced by AGN and thus the identifications of WR features in these galaxies (especially the BPT-AGN) are suspect.  Moreover, WR stars do not exhibit strong, if any, hydrogen lines \citep[e.g.][]{schaerervacca1998, crowther2007,crowtherwalborn2011}.  Therefore, we do not consider WR stars a likely possibility for producing the observed broad \ha\ in the broad-line AGN candidates.  
LBVs are sources of Balmer emission and can exhibit broad \ha\ with velocities as large as $\sim 1500$ km s$^{-1}$, yet they are generally less luminous than Type II SNe \citep{smithetal2011} and we therefore consider them a less likely source of possible contamination in the broad-line sample.  Nevertheless, like SNe, LBVs are transient events that we can eliminate from our broad-line AGN candidates with our follow-up spectroscopic campaign. 

In the spectra of some broad-line AGN candidates, the broad \ha\ components are rather offset from the narrow components.  If a given broad \ha\ line is indeed from gas orbiting a massive BH, this may suggest the AGN is physically offset from the center of the galaxy (perhaps due to a relatively shallow potential well in a low-mass galaxy).  Alternatively, the broad-line could be due to a SNe offset from the galaxy center, which is more likely for objects in the star-forming region of the BPT diagrams with exceptionally broad Ha (e.g, object C).  Our follow-up observations will help discern the origin of the offset lines.

\subsection{Black Hole Masses} 
 
For the objects with broad \ha\ emission, we can estimate indirect BH masses using scaling relations.  If the BLR gas is virialized, for which there is some evidence \citep{petersonetal2004}, then we can use the BLR kinematics as a dynamical tracer of the BH mass $(M_{\rm BH} \propto R \Delta V^2/G)$.  The average gas velocity is inferred from the emission-line width. The BLR size has been measured only for $\sim 50$ active galaxies using reverberation mapping \citep[e.g.,][]{petersonetal2004,bentzetal2009lamp,denneyetal2010,barthetal2011,bentzetal2013}, in which the time lag between continuum and BLR variability establishes a size scale for the BLR.  In general, the geometry of the BLR (and the proportionality constant) is not known, and so the virial product alone ($R \Delta V^2 /G$) does not establish the absolute BH mass.  Instead, the ensemble of reverberation-based BH masses are calibrated using the \msigma\ relation \citep{gebhardtetal2000b,ferrareseetal2001,nelsonetal2004,onkenetal2004,greeneho2006msig,parketal2012,grieretal2013}, although for some caveats
see \citet{greeneetal2010}.  

For all other broad-line AGNs, the BLR size is unknown, and is indirectly estimated using a correlation between AGN luminosity and BLR size (the radius-luminosity relation) that is measured using \hbeta\ time-lag measurements from the reverberation-mapped sample \citep[e.g.,][]{kaspietal2005,bentzetal2009rl,bentzetal2013}.  Since the AGN continuum luminosity is virtually impossible to measure in these low-luminosity galaxies with ongoing star formation, we use the broad Balmer line luminosity as a proxy for the AGN luminosity \citep[e.g.,][]{yeeetal1980}.  

We follow the approach outlined in \citet{greeneho2005cal} to estimate virial BH masses using the FWHM and luminosity of broad \ha, but with the modified radius-luminosity relationship of \citet{bentzetal2013}.  Using the results from the ``clean" fit in \citet{bentzetal2013}, the $R_{\rm BLR}-L$ relationship is given by

\begin{equation}
{\rm log}\left({R_{\rm BLR} \over {\rm lt-days}}\right) = 1.555 + 0.542~{\rm log}\left({L_{5100} \over 10^{44}~{\rm erg~s}^{-1}}\right).
\end{equation}

\noindent
\citet{greeneho2005cal} present well-defined empirical correlations between broad \ha\ luminosity and continuum luminosity, $L_{5100}$,
and the line-widths (FWHM) of \ha\ and \hbeta:

\begin{equation}
L_{\rm H\alpha} = 5.25 \times 10^{42} \left({L_{5100} \over 10^{44}~{\rm erg~s}^{-1}}\right)^{1.157} ~{\rm erg~s^{-1}}
\label{eqn:ha5100}
\end{equation}

\begin{equation}
{\rm FWHM_{H\beta}} = 1.07 \times 10^3 \left({ {\rm FWHM_{H\alpha}}  \over 10^3~{\rm km~s^{-1}}  }\right)^{1.03} ~{\rm km~s^{-1}}.
\end{equation}

\noindent
Inserting the previous three equations into the virial relationship:

\begin{equation}
M_{\rm BH} = \epsilon \left({R_{\rm BLR} {\rm FWHM_{H\beta}}^2 \over G}\right),
\end{equation}

\noindent
gives 

\begin{eqnarray}
{\rm log} \left({M_{\rm BH} \over M_\odot}\right) = {\rm log}~\epsilon + 6.57 + 0.47~{\rm log} \left({L_{\rm H_\alpha} \over 10^{42}~{\rm erg~s^{-1}} }\right) \\
\nonumber
+ 2.06~{\rm log} \left({\rm FWHM_{H\alpha} \over 10^{3}~{\rm km~s^{-1}} }\right).
\label{eqn:mbh}
\end{eqnarray}

\noindent
Several values of the scale factor, $\epsilon$, are found in the literature\footnote{We follow \citet{onkenetal2004} and denote the scale factor as $\epsilon$ when using the FWHM of the line, as opposed to $f$ when using the second moment of the line profile, $\sigma$.}, spanning a range of $\sim 0.75-1.4$ \citep[e.g.,][]{onkenetal2004,greeneho2007,grieretal2013}.  Here we assume $\epsilon = 1$.

Table \ref{tab:blagncands} lists the broad \ha\ luminosities, widths, and virial BH masses for our sample of broad-line AGN candidates.  The most secure broad-line AGN candidates (i.e.\ those that lie in the AGN or composite region of the BPT diagram) have broad \ha\ luminosities of $\sim 10^{39}-10^{40}$ erg s$^{-1}$, with the exception of the nearby dwarf Seyfert NGC~4395 that has a broad \ha\ luminosity of $\sim 10^{38}$ erg s$^{-1}$.  For these sources, the widths (FWHM) of the broad \ha\ components span a range of $\sim 600-1600$ km s$^{-1}$.  The virial BH masses calculated using Equation \ref{eqn:mbh} are in the range $\sim 10^5-10^6$ with a median of $\sim 2 \times 10^5$~\msun.  For comparison, the median BH masses in the \citet{greeneho2007} and \citet{dongetal2012} samples are $\sim 1 \times 10^6$~\msun.  The less secure broad-line sources (in the star-forming region of the BPT diagram) have broad \ha\ luminosities comparable to the more secure broad-line AGN candidates.  However, the widths reach values of a few thousand km s$^{-1}$, leading to anomalously large BH mass estimates and possibly indicating a different origin for the broad \ha\ emission at least in the most extreme cases (Figure \ref{fig:histograms}a, also see Section \ref{sec:broad}).  

\begin{deluxetable}{cccc}
\tabletypesize{\tiny}
\tablecaption{Broad-line AGN Candidates: BH Masses}
\tablewidth{0pt}
\tablehead{
\colhead{ID} & \colhead{log $L({\rm H}\alpha)_b$} & \colhead{FWHM(H$\alpha)_b$} & \colhead{log $M_{\rm BH}$}  \\
\colhead{(1)} & \colhead{(2)} & \colhead{(3)} & \colhead{(4)} }
\startdata
\hline
\multicolumn{4}{c}{\it BPT AGNs} \\
\hline
1  & 39.38 & 1577 & 5.7 \\
9  & 40.15 & 703 & 5.4 \\
11  & 39.41 & 636 & 4.9 \\
20  & 40.13 & 1526 & 6.1 \\
21  & 38.15 & 1288 & 5.0 \\
32  & 39.73 & 747 & 5.2 \\
\hline
\multicolumn{4}{c}{\it BPT Composites} \\
\hline
48  & 39.67 & 894 & 5.4 \\
119  & 40.16 & 1043 & 5.7 \\
123  & 39.82 & 634 & 5.1 \\
127  & 39.45 & 792 & 5.2 \\
\hline
\multicolumn{4}{c}{\it BPT Star-Forming} \\
\hline
A  & 39.52 & 1782 & 5.9 \\
B  & 40.67 & 1245 & 6.1 \\
C  & 40.10 & 3690 & 6.8 \\
D  & 39.56 & 4124 & 6.7 \\
E  & 39.26 & 935 & 5.2 \\
F  & 40.09 & 598 & 5.2 \\
G  & 39.56 & 2027 & 6.1 \\
H  & 39.97 & 3014 & 6.6 \\
I  & 39.39 & 994 & 5.3 \\
J  & 39.99 & 1521 & 6.0 \\
K  & 39.58 & 774 & 5.2 \\
L  & 39.51 & 3126 & 6.4 \\
M  & 39.75 & 3563 & 6.6 \\
N  & 40.42 & 645 & 5.4 \\
O  & 38.88 & 1563 & 5.5 \\
\enddata
\tablecomments{Col.(1): Identification assigned in this paper.
Col.(2): Luminosity of the broad component of H$\alpha$ in units of erg s$^{-1}$.
Col.(3): FWHM of the broad component of H$\alpha$, corrected for instrumental resolution, in units
of km s$^{-1}$. Col.(4): Virial mass estimate of the BH in units of M$_\odot$, assuming the broad H$\alpha$ emission
is due to accretion.  As described in the text, the origin of the broad emission for the BPT star-forming galaxies is unclear at present.}
\label{tab:blagncands}
\end{deluxetable}

The virial BH mass measurements for the broad-line AGN candidates are obviously extremely indirect, and are subject to a number of systematic uncertainties.  For one thing, the BLR geometry clearly varies considerably from object to object \citep{kollatschny2003,bentzetal2009lamp,denneyetal2010,barthetal2011}, making a single geometric scaling factor suspect.  Whether the geometry depends on fundamental parameters of the BH remains uncertain \citep{collinetal2006}.  Then there is the danger that we are not measuring the velocities at the same radius as probed with reverberation mapping \citep{krolik2001}.  At present, there simply are not enough measurements to search for such systematics directly, although progress with velocity-resolved reverberation mapping is promising \citep{gaskell1988,kollatschny2003,bentzetal2008lamp,denneyetal2009b,barthetal2011}.  Finally, reverberation mapping has been achieved in only a few low-mass AGNs \citep{petersonetal2005,rafteretal2011} and we therefore strongly caution that the BH masses presented here are based on an extrapolation from more luminous AGNs powered by more massive BHs and living in very different environments.

\subsection{Broad H$\alpha$ Detection Limits}

For the BPT AGN and composites that {\it do not} have detectable broad \ha\ emission in their spectra, we determine upper limits on any potential broad \ha\ flux.  We add fake broad \ha\ components represented by Gaussians to the emission-line spectra and measure the emission lines with the code described in Section \ref{sec:linefit}.  The input FWHM is held fixed at 500 km s$^{-1}$ to match our selection criteria and we incrementally increase the height of the Gaussian until our code flags the source as having a broad \ha\ component.  The minimum detectable flux is $\sim 10^{-15}$~\fluxunits\ within a factor of $\sim$3 for nearly all of the sources.  This corresponds to a broad \ha\ luminosity of $\sim 2 \times 10^{39}$~erg s$^{-1}$ at a median redshift of $z \sim 0.03$.  We note that all of the BPT AGN and composites with detected broad \ha\ are above this threshhold, with the exception of NGC~4395 that is only $\sim$4 Mpc away.

We estimate the bolometric luminosity of an AGN corresponding to our minimum detectable broad \ha\ luminosity using the conversion between $L({\rm H\alpha})$ and $L({\rm 5100\AA})$ in Equation \ref{eqn:ha5100} \citep{greeneho2005cal} and $L_{\rm bol}=10.3L({\rm 5100\AA})$ \citep{richardsetal2006}.  We caution that these relationships come from studies of more luminous Seyfert 1 galaxies and quasars.  Nevertheless, we use these relationships and estimate that an AGN must have a minimum bolometric luminosity of $L_{\rm bol} \gtrsim 10^{42}$ erg s$^{-1}$ for us to detect broad \ha\ emission (modulo obscuration from dust).

In principle, we could detect broad \ha\ from a BH with a mass as low as $\sim 8 \times 10^3$~\msun\ {\it if} it was accreting at its Eddington limit, where $L_{\rm Edd}=1.3 \times 10^{38}~(M_{\rm BH}/$\msun) erg s$^{-1}$.  Since maximally accreting BHs are very rare \citep{schulzewisotzki2010}, it is not surprising that we do not detect such low-mass BHs (if they exist) in the limited volume of our sample ($z < 0.055$).  In other words, we cannot rule out the existence of $\sim 10^4$~\msun\ BHs.  We just cannot detect them via broad \ha\ if they are radiating much below their Eddington luminosity.  A $10^5$~\msun\ BH, on the other hand, only needs to be radiating at $\sim$8\% of its Eddington luminosity to produce detectable broad \ha\ within our survey volume.  

\section{Host Galaxies}

\begin{figure*}[!t]
\begin{center}
\hspace{-1cm}
{\includegraphics[width=6.5in]{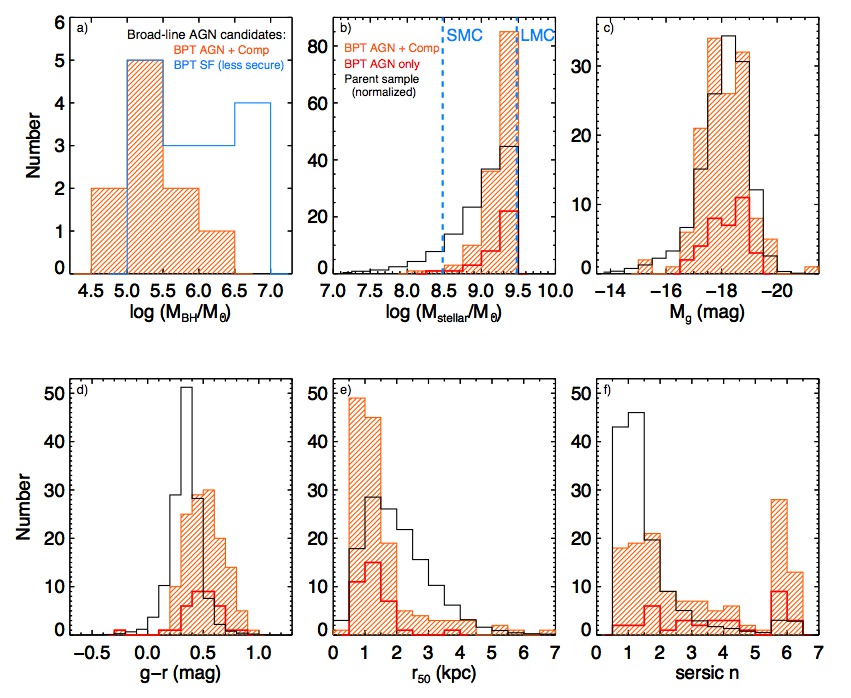}}
\end{center}
\caption{\footnotesize {\it Panel a):} Virial BH mass distribution for broad-line AGN candidates.  BPT AGNs and composites are shown in orange.  BPT star-forming galaxies, for which the origin of the broad-line emission is somewhat ambiguous, are shown in blue.
{\it Panels b--f):} Distributions of host galaxy properties provided in the NSA for the 136 galaxies with narrow-line ionization signatures of BH accretion.  BPT AGNs and composites are shown in the orange hashed histograms, and the distributions of BPT-AGNs only are shown in red.  Our parent sample of dwarf emission-line galaxies is shown in black, normalized to the number of galaxies in the orange histogram (136 BPT AGN + composites).  We show the distributions of galaxy stellar masses (with stellar masses of the Magellanic Clouds indicated in blue), total absolute $g$-band magnitude, $g-r$ color, Petrosian 50\% light radius, and S{\'e}rsic index.  All magnitudes are $K$-corrected to rest-frame values using \texttt{kcorrect v4\_2} and corrected for foreground Galactic extinction.  NSA values have been modified assuming $h=0.73$.}
\label{fig:histograms}
\end{figure*}

The NSA, from which we have drawn our parent sample of dwarfs (Section \ref{sec:sample}), provides a number of galaxy parameters.  In Figure \ref{fig:histograms}b--f, we plot the distributions of various properties for the 136 galaxies that exhibit ionization signatures of BH accretion in their spectra (i.e., those galaxies with narrow-line ratios falling in the AGN and composite region of the BPT diagram).  We do not include the 15 broad-line AGN candidates falling in the star-forming region of the BPT diagram for the reasons discussed in Section \ref{sec:broad}.  For comparison, we also show the distributions for our parent sample of emission-line galaxies normalized to the same number sources (136), as well as the 35 BPT-AGNs alone.  

By design, the galaxies in our sample have stellar masses $M_{\star} \lesssim 3 \times 10^9$ M$_\odot$, which is approximately the stellar mass of the LMC \citep{vandermareletal2002}.  While active BHs are preferentially found in more massive galaxies within our parent sample, we do detect active BHs in galaxies 10 times less massive than our threshold with stellar masses comparable to the Small Magellanic Cloud \citep[SMC, $M_{\star} \sim 3 \times 10^8$ M$_\odot$;][]{stanimirovicetal2004}.  Stellar masses in the NSA are derived from \texttt{kcorrect} \citep{blantonroweis2007} using Galaxy Evolution Explorer (GALEX) UV and SDSS $ugriz$ bands combined, and while they do not account for any possible AGN contribution, we do not expect the active BHs in our sample to significantly impact stellar mass estimates of the host galaxies.  The use of many broadband filters should help mitigate the contribution from strong emission lines, and the vast majority of sources are type 2 AGN that are known to contribute very little continuum emission \citep[e.g.,][]{schmittetal1999}.  For the small fraction of type 1 AGN candidates, the host galaxy stellar masses could in principle be artificially elevated by a boost in flux from the AGN continuum.  However, {\it HST} $I$-band imaging of two of our broad-line AGN candidates \citep[IDs 20 and 123;][]{jiangetal2011} that are also in the sample of low-mass BHs presented by \citet{greeneho2007} reveals that the AGNs in these galaxies contribute $\lesssim 10\%$ of the total flux.

The active galaxies in our sample have a total (host + AGN) median absolute $g$-band magnitude of $\langle M_g \rangle = -18.1$ mag, which is comparable to the LMC \citep[$M_g^{\rm LMC} \sim -18.2$ mag; ][]{tollerudetal2011} and $\sim$1--2 magnitudes fainter than previous samples of low-mass AGN.  Correcting the median absolute magnitudes to our adopted cosmology ($h=0.73$), the \citet{barthetal2008a} sample of low-mass Seyfert 2 galaxies has $\langle M_g \rangle = -19.0$, and the low-mass Seyfert 1 samples of \citet{greeneho2007} and \citet{dongetal2012} have $\langle M_g \rangle = -19.4$ mag and $\langle M_g \rangle = -20.3$ mag, respectively, after removing the AGN contribution. 

The colors of our sample of dwarfs hosting active BHs tend to be redder compared to our parent sample of dwarfs.  The distribution of colors remains essentially the same for the parent sample even when applying a mass cut of log $M_\star > 9.25$, and therefore the color difference between the active sample and parent sample is not a mass effect.  The $A_V$'s of the active galaxies derived from our continuum and absorption line fits are similar to the parent sample ($\sim$0 to 1 with a median of $\sim 0.3$), suggesting the redder colors are not due to differences in dust properties.  More likely, the color difference is a selection effect such that the optical diagnostics we are using are not sensitive to accreting BHs in blue galaxies with ongoing star-formation that dominates the emission-line spectra. 

In addition to being low-mass, the active galaxies in our sample are physically small, with typical half-light radii $r_{50} \lesssim 2$~kpc.  The distribution of BPT-AGN and composites is skewed towards smaller sizes relative to our parent sample of dwarf galaxies.  A similar trend is seen in the distribution of 90\%-light radii, suggesting the galaxies hosting active BHs are indeed preferentially compact and this result is not due to the presence of a bright central point source.  

\begin{figure*}[!t]
\vspace{0.9cm}
\epsscale{1}
\plotone{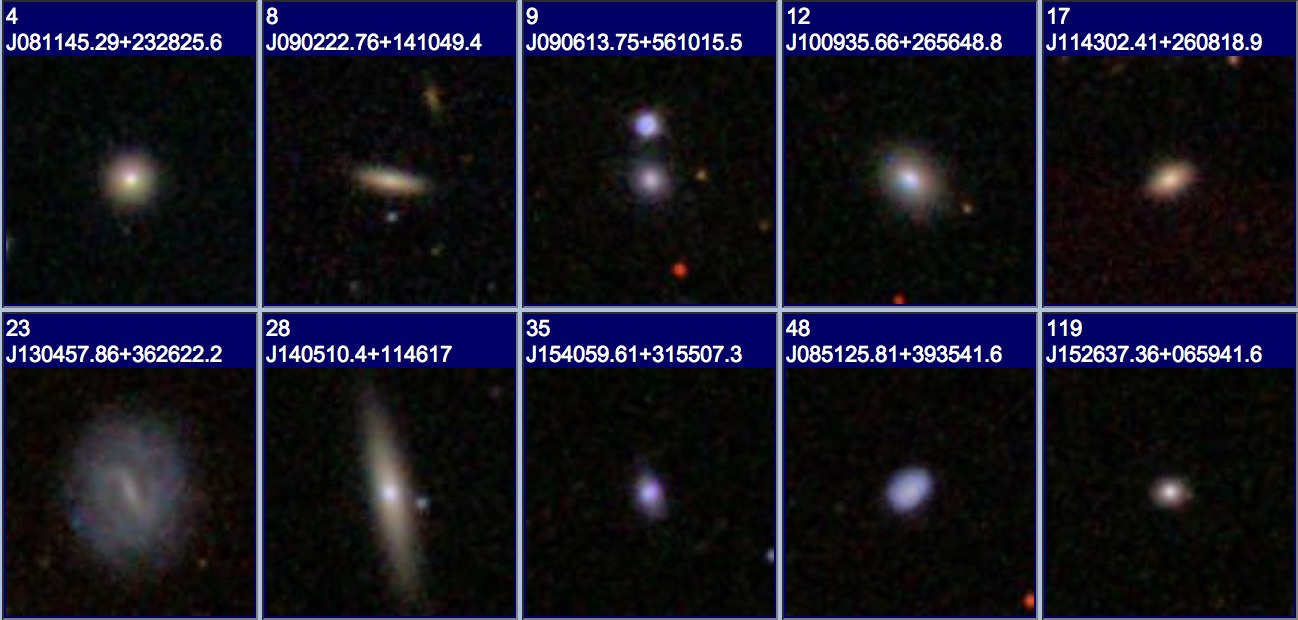}
\caption{\footnotesize A selection of galaxies from our sample of dwarfs hosting active massive BHs.  The SDSS color composite
images have a size of $50\arcsec \times 50\arcsec$.  The identification numbers assigned in this work are in the upper
left corners of the images, with the SDSS name below.}\label{fig:sdssims}
\end{figure*}

Single-component two-dimensional S{\'e}rsic fits are also provided in the NSA and the galaxies in our sample span a large range of S{\'e}rsic index, $n$.  Relative to our parent sample, a larger fraction of the active galaxies have high-$n$ values, although BHs are also found in low-$n$ disky galaxies.  Figure \ref{fig:sdssims} shows the SDSS images of a selection of galaxies in our sample with BH accretion signatures.  

\section{Demographics}\label{sec:discussion}

The fraction of {\it optically-selected, active BHs} in our parent sample of dwarf galaxies is $\sim$0.5\% (136/25974)\footnote{Including the additional 15 broad-line AGN candidates in the star-forming region of the BPT diagram does not have a significant impact on the active fraction, increasing it to $\sim$0.6\%.}.  However, there are a number of obstacles preventing us from determining the true occupation fraction and BH mass function in this low-mass regime.  First of all, our optical diagnostics are only sensitive to actively accreting BHs, and even at their Eddington limit low-mass BHs are relatively faint.  Furthermore, small galaxies generally have ongoing star formation, gas, and dust that can mask or extinguish the optical signatures of BH accretion.  Therefore, while there may an accreting BH present at the center of a galaxy, the total observed line emission in the SDSS aperture may be dominated by star formation.  Indeed, the SDSS aperture of 3\arcsec\ is comparable to the median half-light radius of our dwarf galaxy sample.  Even without significant ongoing star formation, AGN signatures may be heavily diluted by host galaxy light such that the emission lines are effectively hidden \citep{moranetal2002}.  Additionally, low-metallicty AGN, which may be expected in lower-mass galaxies, can fall (and hide) in the upper left region of the star-forming plume of galaxies in the \OIII/\hbeta\ versus \NII/\ha\ diagnostic diagram \citep{grovesetal2006}.  Therefore, while we can identify bona-fide AGNs based on their location in the BPT diagram, the selection of massive BHs is likely highly incomplete.  Even if we understand our incompleteness from these effects, to derive a true space density requires that we know the distribution of Eddington ratios in these low-mass systems as compared to more massive galaxies \citep[e.g.,][]{heckmanetal2004,galloetal2010,airdetal2012} where we believe the occupation fraction is close to unity.  It is interesting to note, however, that our active fraction is quite similar to that of $\sim 10^7$~\msun\ BHs radiating at $\sim$10\% of their Eddington limit \citep{heckmanetal2004,greeneho2007b}.  

Using broad emission lines to identify AGN in dwarf galaxies poses a different set of problems.  The broad-line signature is weaker for low-mass BHs and can be difficult to detect in galaxy-dominated spectra.  There is also the possibility that the broad-line region disappears altogether below some critical luminosity or Eddington ratio
\citep[e.g.,][]{nicastro2000,laor2003,elitzurho2009,trumpetal2011,marinuccietal2012}.  There are a number of candidate ``true'' Type 2 AGNs \citep[e.g.,][]{tran2003,bianchietal2008}
that show no sign of a broad-line region in direct or polarized light, and no clear signs of obscuration in X-rays.  It is thus possible that the Type 1 fraction drops towards lower luminosity, which could add significant complications in attempting to use these AGN as a tracer of the demographics of BHs in dwarf galaxies.

\section{Conclusions}

Using optical spectroscopy from the SDSS, we have systematically assembled the largest sample of dwarf galaxies ($10^{8.5} \lesssim M_{\star} \lesssim 10^{9.5}$~\msun) hosting massive BHs to date.  These dwarf galaxies have stellar masses comparable to the Magellanic Clouds and contain some of the least-massive supermassive BHs known.  Contrary to common lore, low-mass, physically small dwarf galaxies can indeed form massive BHs.  

We find photoionization signatures of BH accretion in 136 galaxies using the narrow-line \OIII/\hbeta\ versus \NII/\ha\ diagram as our primary diagnostic.  Of these, 35 have AGN-dominated spectra and 101 have composite spectra suggesting ionization from both an AGN and massive stars.  For the small fraction of these active galaxies with detectable broad \ha\ emission, we estimate a median virial BH mass of $M_{\rm BH} \sim 2 \times 10^{5}$~\msun.  Our sensitivity to broad \ha\ emission limits our ability to detect broad-line AGN with BH masses much below $\sim 10^5$~\msun\ radiating at less than $\sim$10\% of their Eddington luminosity.  We find broad \ha\ in an additional 15 galaxies, yet their spectra exhibit narrow-line ratios consistent with star-forming galaxies.  We caution that at these low-luminosities and low-metallicities, particularly for galaxies with high star formation rates, we are susceptible to contamination from stellar processes.

Ultimately, we need a complete census of massive BHs in dwarf galaxies to place stringent constraints on theories for the formation of supermassive BH seeds.  While optical diagnostics certainly have a role to play, we need to move towards using alternative search techniques and observations at other wavelengths to make further progress \citep[e.g., radio and X-ray;][]{reinesetal2011,reinesdeller2012,galloetal2010,milleretal2012,kamizasaetal2012}.

\acknowledgments

We are grateful to the entire SDSS collaboration for providing the data that made this work possible, to Michael Blanton and all those involved in creating the NASA-Sloan Atlas, and to Craig Markwardt for making his MPFIT code publicly available.  We thank the referee for a very helpful review that improved the paper.  A.E.R. appreciates helpful discussions with Mark Whittle, Jong-Hak Woo and Marta Volonteri.  Support for A.E.R. was provided by NASA through the Einstein Fellowship Program, grant PF1-120086.  J.E.G. is partially supported by an Alfred P. Sloan fellowship.  

\newpage

\appendix

\section{Spectra of the Broad-line AGN Candidates}

\begin{figure*}[!h]
\epsscale{1.1}
\plotone{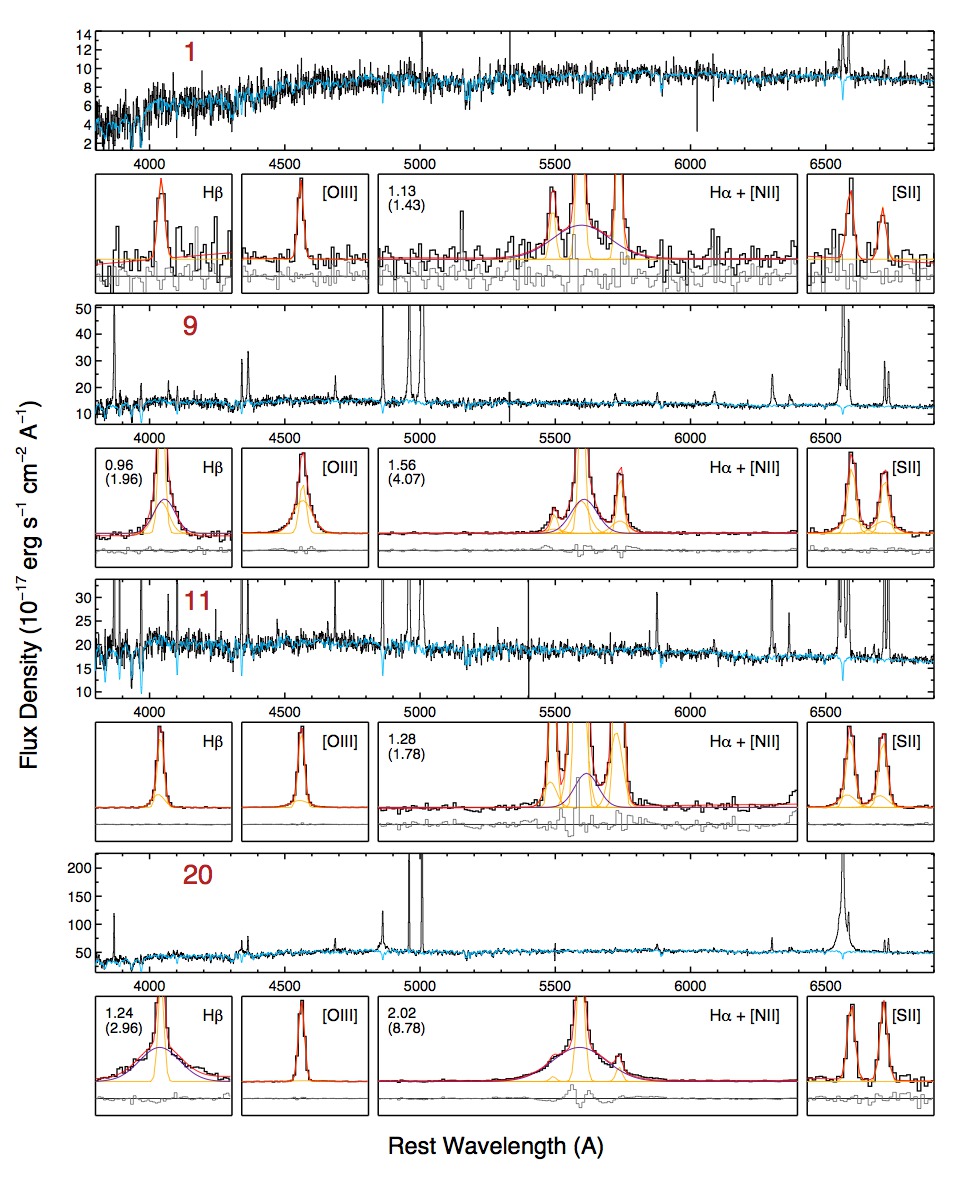}
\caption{\footnotesize Same as Figure \ref{fig:broadex}.}
\label{fig:blagn_spec1}
\end{figure*}

\newpage

\begin{figure*}[!h]
\epsscale{1.1}
\plotone{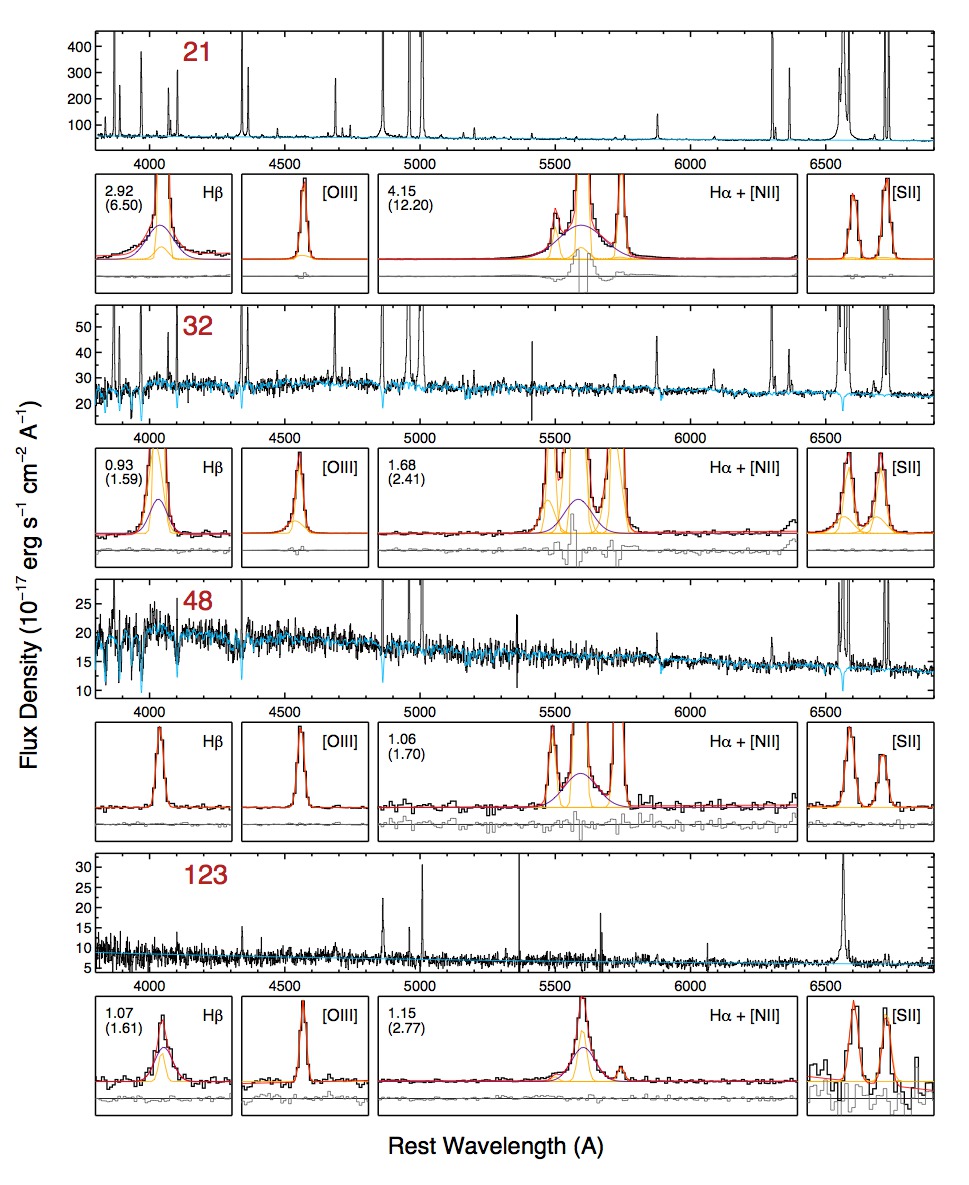}
\caption{\footnotesize Same as Figure \ref{fig:broadex}.  Source 119 is shown in Figure \ref{fig:broadex}.}
\label{fig:blagn_spec2}
\end{figure*}

\newpage

\begin{figure*}[!h]
\epsscale{1.1}
\plotone{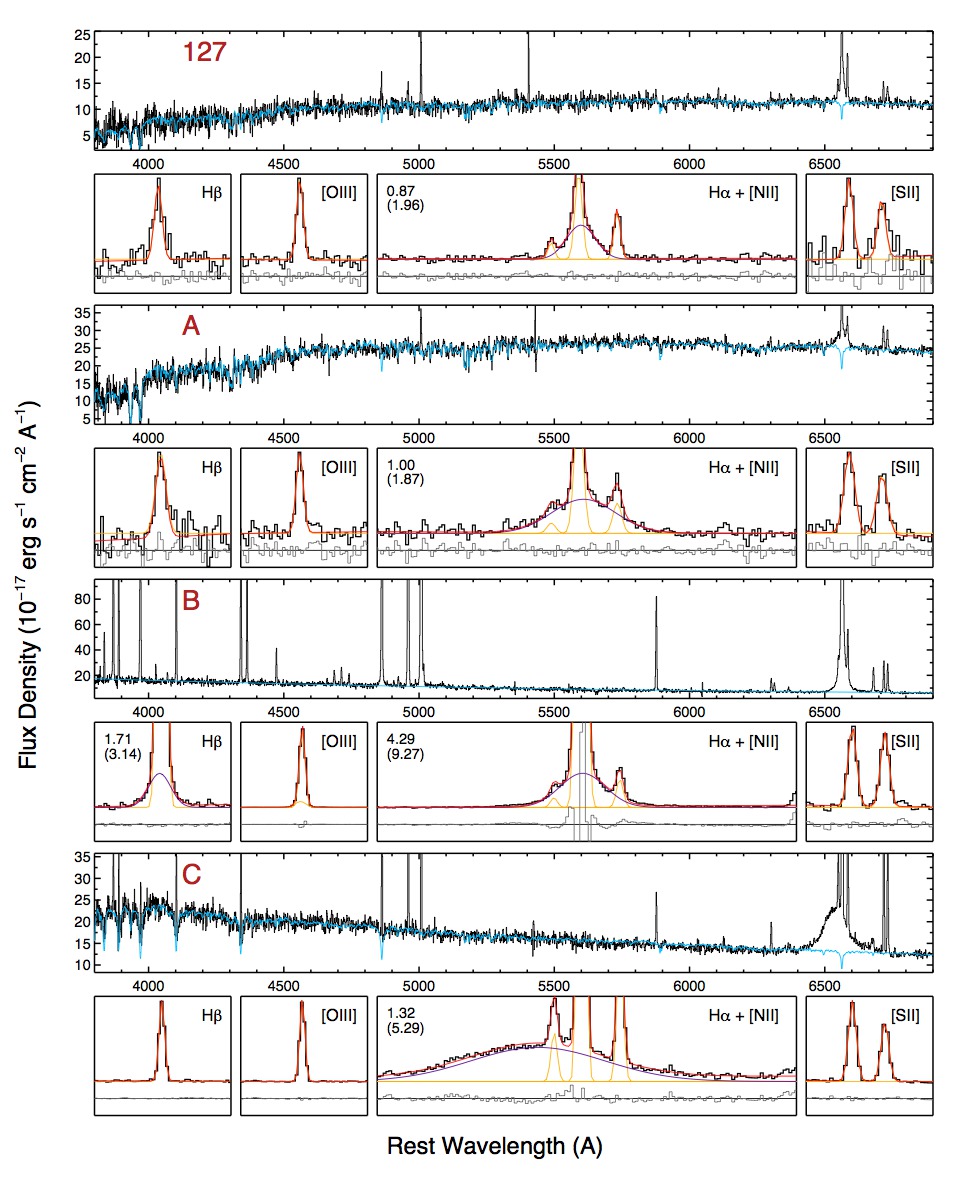}
\caption{\footnotesize Same as Figure \ref{fig:broadex}.}
\label{fig:blagn_spec3}
\end{figure*}

\newpage

\begin{figure*}[!h]
\epsscale{1.1}
\plotone{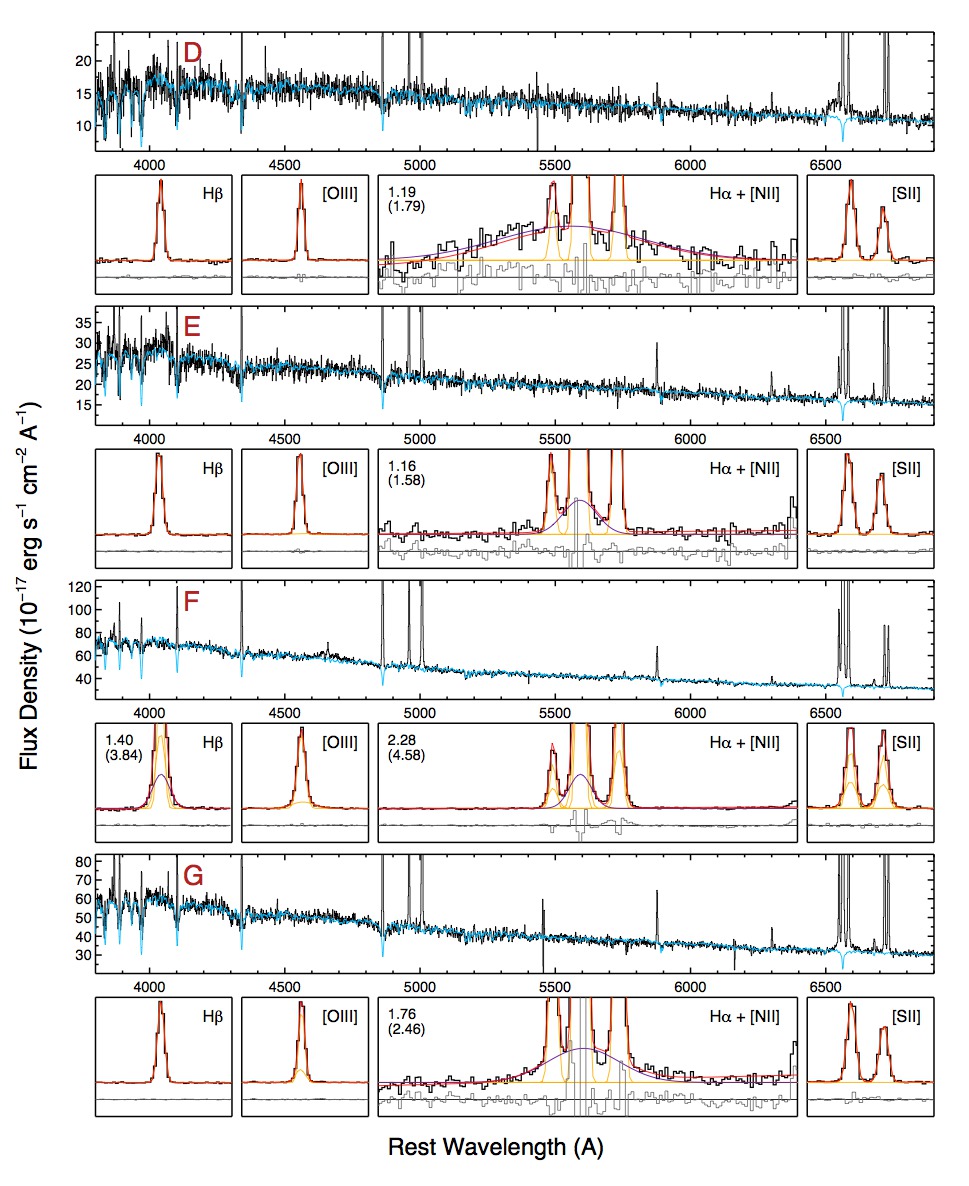}
\caption{\footnotesize Same as Figure \ref{fig:broadex}.}
\label{fig:blagn_spec4}
\end{figure*}

\newpage

\begin{figure*}[!h]
\epsscale{1.1}
\plotone{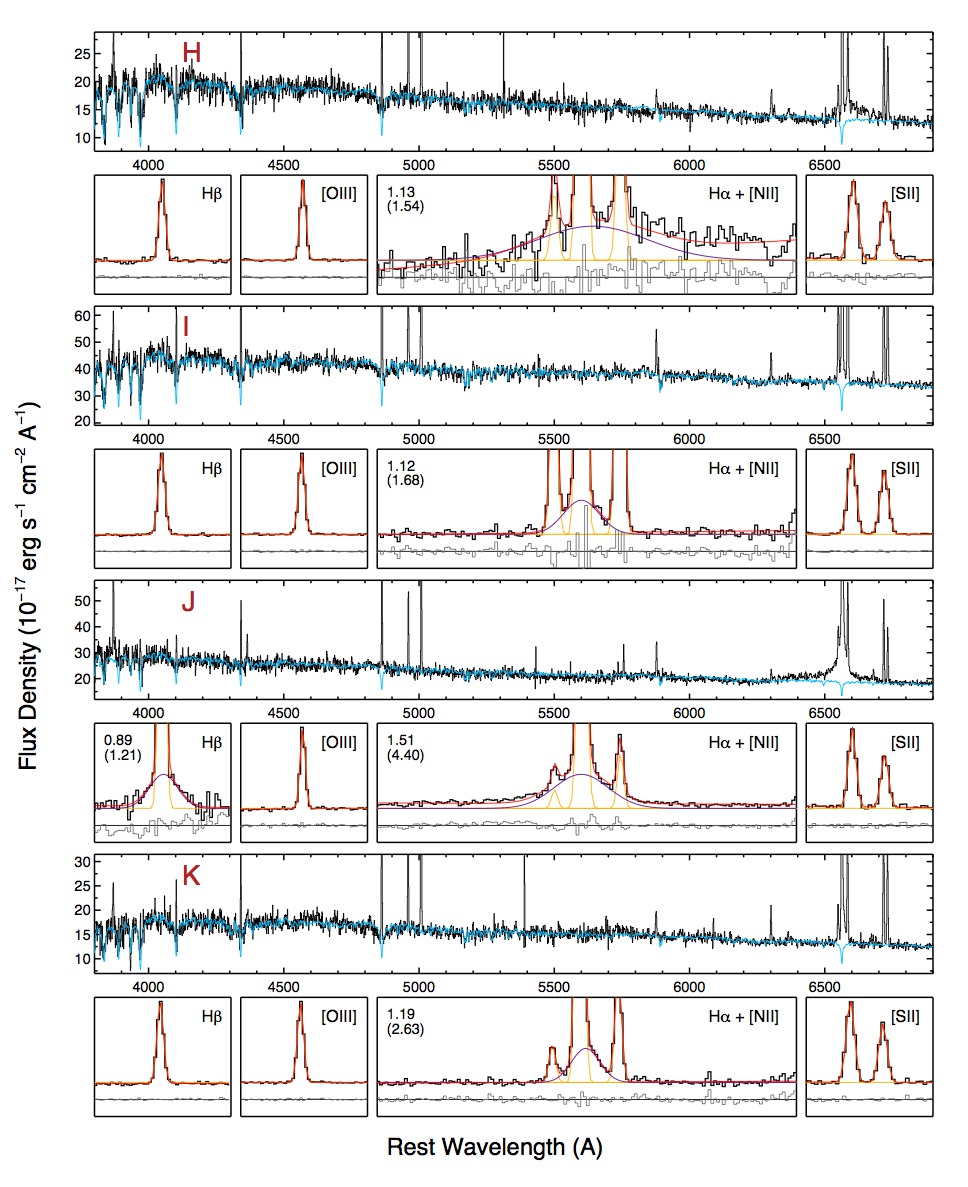}
\caption{\footnotesize Same as Figure \ref{fig:broadex}.}
\label{fig:blagn_spec5}
\end{figure*}

\newpage

\begin{figure*}[!h]
\epsscale{1.1}
\plotone{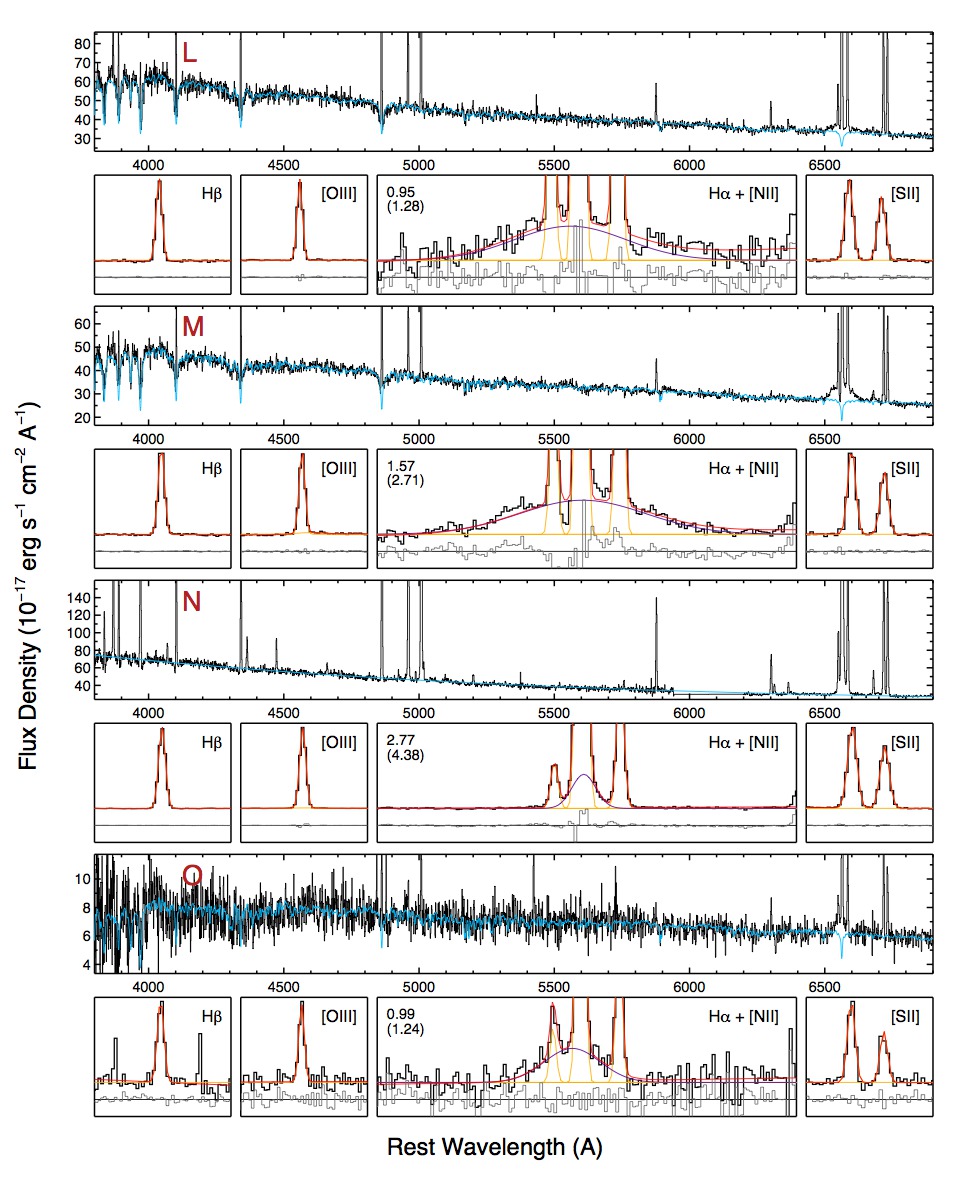}
\caption{\footnotesize Same as Figure \ref{fig:broadex}.}
\label{fig:blagn_spec6}
\end{figure*}

\newpage


\end{document}